\documentclass[superscriptaddress, preprint, amsmath, amssymb, aps,pra, floatfix]{revtex4-1}

\usepackage{graphicx, makecell, multirow}
\usepackage{bm, xurl, setspace, booktabs}

\renewcommand{\thetable}{\arabic{table}}

\usepackage{xcolor}
\definecolor{myorange}{RGB}{230, 81, 0}
\definecolor{mypurple}{RGB}{170, 0, 255}
\definecolor{mygreen}{RGB}{76, 175, 80}
\definecolor{myblue}{RGB}{33,150,243}
\definecolor{verylightgray}{gray}{0.6}

\begin{document}

\title{LLMs Can Infer Political Alignment from Online Conversations}

\author{Byunghwee Lee}
\thanks{These authors contributed equally to this work.}
\affiliation{School of Data Science, University of Virginia, Charlottesville, Virginia, USA, 22903}
\affiliation{Center for Complex Networks and Systems Research,\\Luddy School of Informatics, Computing, and Engineering\\ Indiana University, Bloomington, Indiana, USA, 47408}

\author{Sangyeon Kim}
\thanks{These authors contributed equally to this work.}
\affiliation{Center for Complex Networks and Systems Research,\\Luddy School of Informatics, Computing, and Engineering\\ Indiana University, Bloomington, Indiana, USA, 47408}
\affiliation{Division of Communication and Media,\\Ewha Womans University, Seoul, Republic of Korea, 03760}

\author{Filippo Menczer}
\affiliation{Center for Complex Networks and Systems Research,\\Luddy School of Informatics, Computing, and Engineering\\ Indiana University, Bloomington, Indiana, USA, 47408}

\author{Yong-Yeol Ahn}
\email{yyahn@virginia.edu}
\thanks{Corresponding author}
\affiliation{School of Data Science, University of Virginia, Charlottesville, Virginia, USA, 22903}
\affiliation{Center for Complex Networks and Systems Research,\\Luddy School of Informatics, Computing, and Engineering\\ Indiana University, Bloomington, Indiana, USA, 47408}

\author{Haewoon Kwak}
\email{hwkwak@iu.edu}
\thanks{Corresponding author}
\affiliation{Center for Complex Networks and Systems Research,\\Luddy School of Informatics, Computing, and Engineering\\ Indiana University, Bloomington, Indiana, USA, 47408}

\author{Jisun An}
\email{jisunan@iu.edu}
\thanks{Corresponding author}
\affiliation{Center for Complex Networks and Systems Research,\\Luddy School of Informatics, Computing, and Engineering\\ Indiana University, Bloomington, Indiana, USA, 47408}

\begin{abstract}
Due to the correlational structure in our traits such as identities, cultures, and political attitudes, seemingly innocuous preferences like following a band or using a specific slang can reveal private traits. This possibility, especially when combined with massive, public social data and advanced computational methods, poses a fundamental privacy risk. As our data exposure online and the rapid advancement of AI are increasing the risk of misuse, it is critical to understand the capacity of large language models (LLMs) to exploit such potential. Here, using online discussions on Debate.org and Reddit, we show that LLMs can reliably infer hidden political alignment, significantly outperforming traditional machine learning models. Prediction accuracy further improves as we aggregate multiple text-level inferences into a user-level prediction, and as we use more politics-adjacent domains. We demonstrate that LLMs leverage words that are highly predictive of political alignment while not being explicitly political. Our findings underscore the capacity and risks of LLMs for exploiting socio-cultural correlates.
\end{abstract}

\maketitle


\section{Introduction}

Publicly visible preferences, such as driving a pickup truck vs. an electric vehicle, choosing brewed tea or latte~\cite{mutz2018real}, or preferring particular musical genres or bands, often correlate with \emph{private} characteristics including age, gender, sexual orientation, religion, personality, and political leaning~\cite{dellaposta2015liberals, hetherington2018prius, mutz2018real, talaifar2025lifestyle}. 
Such correlations may emerge from underlying homophily~\cite{huber2017political}, sorting in social and physical environments~\cite{bourdieu2018distinction, lizardo2006cultural, bishop2009big}, geographical confounds~\cite{rodden2010geographic, martin2020does}, congruence~\cite{lakoff2008metaphors} and stochastic social processes driven by early variation~\cite{macy2019opinion}. 
Individuals form cognitive associations among beliefs and behaviors through observation and peer influence, a process explained by associative diffusion~\cite{goldberg2018beyond} or social contagion models~\cite{aiyappa2024weighted}. 
As a result, even deeply personal, sensitive, and private information can be inferred from publicly available, seemingly innocuous signals, sometimes with surprisingly high accuracy, particularly when sufficient behavioral information is available and models are carefully trained for the inference task~\cite{kosinski2013private, youyou2015computer}.

The capacity to predict individual traits is central to personalization in services like recommendation systems and search engines~\cite{koren2009matrix}. 
Accordingly, a large body of work has analyzed digital traces as a means of uncovering personal characteristics and preferences. 
Facebook Likes~\cite{kosinski2013private, youyou2015computer}, web browsing histories~\cite{goel2012does}, personal website content~\cite{marcus2006personality}, and music listening records~\cite{rentfrow2003re} have all been used to infer demographic and psychological traits. 
Yet, while offering benefits of more useful recommendations, the ability to infer sensitive attributes, including political leanings, creates substantial risks for misuse that extend far beyond what users typically anticipate. 

The risks become more acute when such inference is scaled across millions of users. 
Political micro-targeting, for example, uses personal data to deliver tailored messages to specific voter segments~\cite{borgesius2018online, haenschen2023conditional, votta2024does}. 
The 2018 Cambridge Analytica scandal made this danger visceral to the masses~\cite{hsu_2018, us_senate_2018, brown2020should}, sparking an ongoing debate among scholars and policymakers about the surveillance and manipulation~\cite{tufekci2014engineering, susser2019technology, bakir2020psychological}.

The barrier to inferring personal traits has been further reduced with the emergence of large language models (LLMs). 
Through web-scale pre-training, LLMs natively encode socio-cultural correlates even without the bespoke training data that traditional machine learning (ML) systems required.
Once the exclusive domain of data experts with access to massive training data, personal profiling is now possible with simple prompts. 
LLMs have been shown to infer personality traits not only from brief open-ended narratives~\cite{wright2026assessing}, but also from the content they consume~\cite{simchon2023online}. 
GPT-4 has been shown to generate personality assessments of public figures that align closely with human judgments, even without task-specific training~\cite{cao2024large}. 
Beyond personality, recent studies demonstrate that LLMs can detect political stances, issue framings, and ideological language in text~\cite{marino2024integrating, kristensen2025chatbots, simchon2024persuasive}, indicating their growing capabilities for political inference.

Recent research further reveals that LLMs encode rich internal representations of high-level concepts such as 
linguistic structure~\cite{bricken2023towards} and subjective human perspectives, including sentiment~\cite{tigges2023linear} and humor~\cite{von2024language}. 
LLMs can capture nuanced relational structures among diverse human beliefs~\cite{lee2025semantic}. 
These findings lead us to investigate how well LLMs can detect political alignment from general and public online conversations. 

Here, we investigate the following research questions:
How accurately can LLMs infer a user's political alignment from their posts on online forums? Can we quantify the ease of inference for each topic? What kind of lexical cues are most revealing? 

To address these questions, we evaluate two widely used LLMs, both proprietary and open-source, on their ability to infer political alignment using text data from Debate.org and Reddit. 
We find that both models can reliably identify political alignment even from discourse that is not explicitly political. 
Furthermore, prediction accuracy improves substantially when multiple text-level predictions are aggregated for each user, and when weighted by LLM confidence. 
We also find that text from categories that are thematically related
to politics, or that have substantial user overlap with political discussions, yields better predictions.
Finally, we observe that the models have higher confidence in keywords that are more politically charged, including not only overtly political terms like \textit{abortion} or \textit{taxes} but also cultural references such as \textit{Taylor Swift} or \textit{Tesla} that have become politicized~\cite{gillingham2025musk, jackson2025taylor}.
These high-confidence words contribute disproportionately to accurate inference, suggesting that LLMs indeed systematically encode the socio-cultural correlates and nuances.
In summary, we show that LLMs can natively detect political signals even in general-domain discourse. Our findings reinforce concerns about privacy and the risks of large-scale political micro-targeting and other misuses, raising urgent questions about appropriate ethical and regulatory responses.

\section{Results}

We use datasets from Debate.org (DDO) and Reddit.
First, from DDO dataset~\cite{durmus2019corpus, durmus2018exploring}, we extract 3,511 users who self-identified as Republican ($n=1{,}776$) or Democrat ($n=1{,}735$). 
These users participated as debaters in 18,602 debates, posting 22,265 arguments. These debates were assigned by the initiator to one of 23 pre-defined categories, including ``Politics'' and other topics (e.g., ``Religion,'' ``Science,'' and ``Music''). We refer all topics other than Politics as \emph{general} topics for the purpose of this study (see Methods for details). 

We further extended our analysis to Reddit. Because Reddit users do not explicitly disclose their political alignment, we focused on users whose political alignment can be inferred with high certainty, based on their activity in two major partisan communities, \texttt{r/Conservative} and \texttt{r/democrats}. 
From the two subreddits, we identified users who consistently received positive feedback within their respective communities by measuring mean scores (upvotes minus downvotes) of their comments on submissions ranked as ``popular'' by Reddit at the time of the experiment, ensuring that the selected users were supported by fellow partisans. 
Among the users who met this criterion, we randomly sampled 1,000 users from each subreddit and assigned political alignment labels (`Republican' for r/Conservative and `Democratic' for r/democrats). Full details are provided in Data and Methods. 
To validate this labeling, we conducted a human annotation experiment (Section~\ref{sec:reddit_valid}; Figure~\ref{fig:reddit_validation} in SI), in which five annotators classified user political alignment from sampled comments. The annotators achieved a mean accuracy of 0.85 and a majority-vote accuracy of 0.92, with moderate inter-annotator agreement (Fleiss $\kappa = 0.576$), confirming the reliability of the community-based labeling.

For each user, we retrieved up to 100 of their most recent comments across all subreddits, spanning both political and general-interest domains. 
Because Reddit comments are typically short (88.6\% contain five tokens or fewer), we aggregated comments at the subreddit level: for each user, all comments posted within the same subreddit were concatenated into a single text. Consequently, the number of texts per user corresponds to the number of unique subreddits they participated in.  
To enable comparison with DDO, we then used GPT-4o to classify each subreddit into the same 23 categories defined in the DDO dataset, based on the subreddit descriptions (see Table~\ref{tab:subreddit_examples} and~\ref{tab:subreddit_examples_cont} for representative examples).
The final dataset after pre-processing contains 19,412 texts across 5,818 subreddits from 993 Republican users and 26,548 texts from 999 Democratic users across 6,736 subreddits (see Methods for details on the pre-processing procedure). 

We used two representative LLMs, GPT-4o (\texttt{GPT-4o-2024-08-06}) and  Llama-3.1-8B (\texttt{Llama-3.1-8B-instruct}), to infer political alignments from user texts. 
GPT-4o, a flagship proprietary model from OpenAI, is widely recognized for its leading performance across various language tasks at the time of this study~\cite{openai_gpt4o_2024}. 
In contrast, Llama-3.1-8B~\cite{grattafiori2024llama} is an open-source model with fewer parameters, offering greater accessibility and reproducibility. 
The two models represent complementary approaches to LLM-driven inference, balancing performance and accessibility. 

\subsection{LLM-based political alignment inference}

\begin{figure}
\centering 
\includegraphics[width=0.95\textwidth]{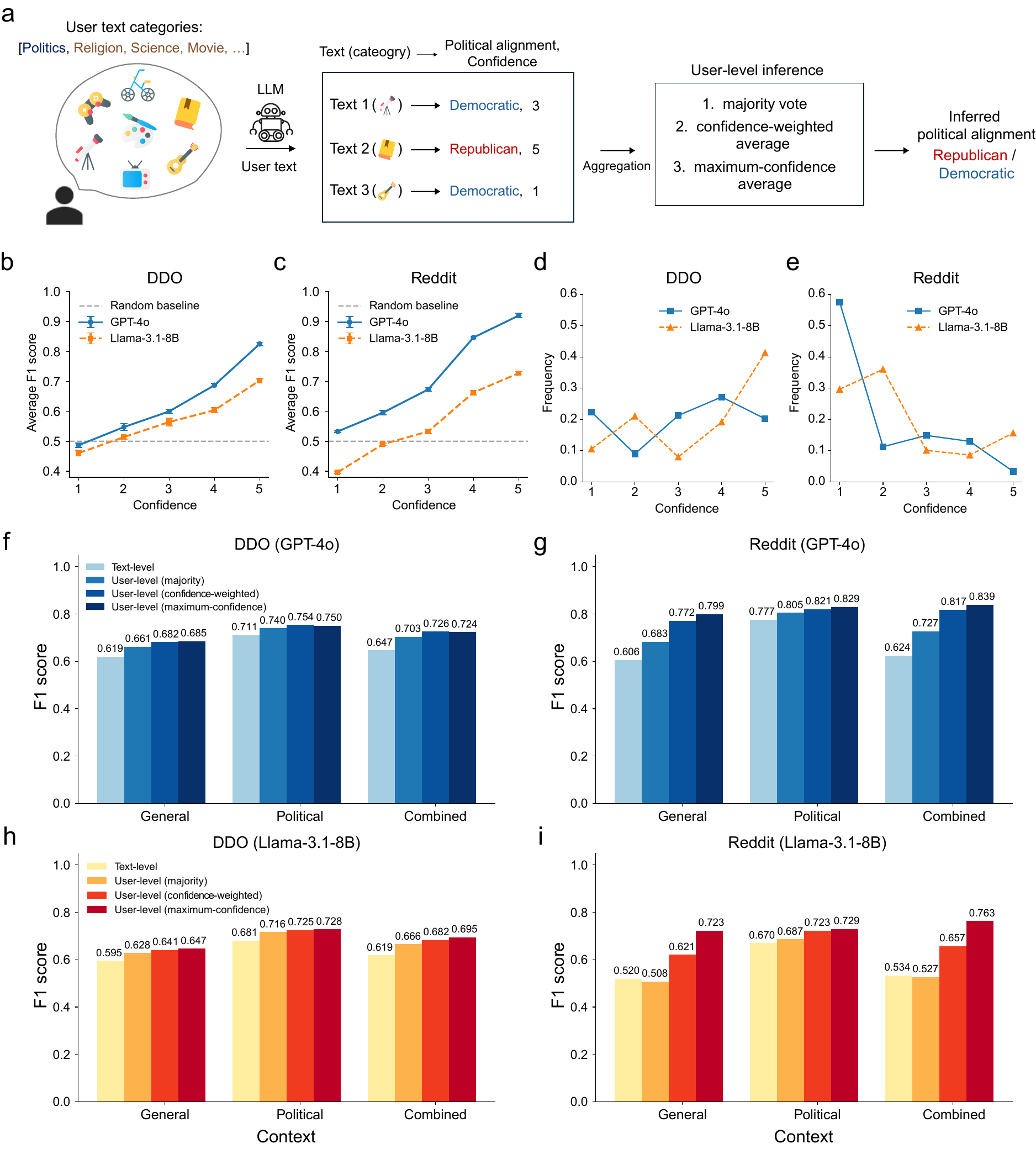}
\caption{\scriptsize
\textbf{LLMs can reliably infer political alignment from general conversations.} (a)~Illustration of the political alignment inference process using LLMs applied to user-generated texts. Text categories can be political (in blue) or general (orange). 
(b-c)~Relationship between average text-level F1 score for the political alignment inference task and LLM confidence scores for (b)~DDO and (c)~Reddit. Error bars represent standard error estimated via bootstrapping (n = 1,000). The average F1 score increases with the LLM's confidence in the text-level inference.
(d-e)~Distribution of confidence for two datasets, (d)~DDO and (e)~Reddit, as reported by the two LLMs. 
(f-i)~Accuracy of LLMs on the political alignment inference task for DDO and Reddit datasets. User-level F1 scores are calculated for three text contexts: general, political, and combined. Results are presented for two LLMs (GPT-4o and Llama-3.1-8B) across three aggregation methods (see main text): (1)~majority vote, (2)~confidence-weighted average, and (3)~maximum-confidence average. Text-level F1 scores are also shown for comparison. Panels are divided by dataset (DDO and Reddit) and model. Across all data and LLMs, user-level aggregation methods that incorporate LLM confidence scores substantially improve inference accuracy. Among the three methods, the maximum-confidence average tends to yield the highest performance, highlighting the effectiveness of leveraging highly confident predictions during inference. All the differences among aggregation methods are statistically significant based on sample paired t-test obtained by bootstrapping ($p<0.01$).}
\label{fig:overview}
\end{figure}

We conduct political alignment inference at both text-level and user-level. In the text-level inference task, we predict a user's political alignment from a single text, which serves as the text-level unit of analysis. For DDO, this unit corresponds to an individual debate argument, whereas for Reddit, it corresponds to a concatenated set of a user's comments within a given subreddit.
During inference, both models were asked to report confidence scores on an integer scale ranging from 1 (least confident) to 5 (most confident), as illustrated in Figure~\ref{fig:overview}a.
(See Figs.~\ref{fig:prompt-gpt-ddo}--\ref{fig:prompt-llama-reddit} for the detailed prompts used for inference with both models across the two datasets.)

At the text-level, both LLMs achieve above-chance performance in inferring political alignment: GPT-4o and Llama-3.1-8B achieve F1 scores of 0.647 and 0.619 on DDO, and 0.624 and 0.534 on Reddit, respectively. We note that in this binary classification task, a random classifier is expected to achieve a macro F1 score of 0.5.
When texts are grouped according to model-reported confidence scores, the average text-level macro F1 score increases with confidence (Figure~\ref{fig:overview}b,c). This trend is consistent across both datasets and models.
For example, the inference accuracy of GPT-4o on the lowest-confidence text group is poor (F1 around 0.5). 
In contrast, the performance on the highest-confidence group reaches F1 scores above 0.8. 
These results indicate that a model's confidence for a given text provides a reliable signal of how accurate is the inferred alignment. Figure~\ref{fig:overview}d,e present the distributions of confidence scores for both models across all analyzed texts. Whereas the DDO dataset shows a somewhat uniform distribution of confidence, the Reddit dataset is skewed toward lower confidence, likely reflecting the more casual and informal nature of Reddit discourse compared to the structured debates in DDO.  

\subsection{User-level political alignment inference}

Having examined inference performance at the text level, we then turn to the user-level prediction by aggregating predictions across multiple texts authored by the same individual. Model-reported confidence can be incorporated into the aggregation process. 
To this end, we evaluate three aggregation methods: (1)~\textit{majority} vote, (2)~\textit{confidence-weighted} average, and (3)~\textit{maximum-confidence} average. 
The majority method serves as the baseline. In this method, the political alignment of a user is inferred by taking the most frequently predicted label across all text-level predictions, without incorporating confidence. 
The confidence-weighted method computes the average alignment score, ranging from 0 (Democratic) to 1 (Republican), by taking the weighted sum of each text-level prediction according to its confidence and then rounding to the nearest integer (0 or 1). 
Finally, the maximum-confidence method applies majority voting to the subset of texts with the highest confidence.

Figure~\ref{fig:overview}f--i compare political alignment inference accuracy across the two LLMs, the two datasets, and text-/user-level classification with the three aggregation methods. 
We report macro F1 scores across three text types---general, political, and combined. 
Incorporating confidence into the user-level inference improves prediction accuracy.
Across the 12 model–dataset–text type combinations, we compare three aggregation methods for user-level inference. In 10 out of these 12 settings, maximum-confidence achieves the highest performance among the aggregation methods, followed by confidence-weighted. 
For example, when inferring the political alignment of DDO users from general text, GPT-4o attains F1 = 0.685 with the maximum-confidence method and F1 = 0.682 with the confidence-weighted method.   
This performance is achieved while relying on only 30\% of user texts (see Table~\ref{tab:maxconf-prop} for the average numbers of texts per user available to the different aggregation methods across model–dataset combinations). 
For political texts, the maximum-confidence and confidence-weighted methods reach F1 scores around 0.750. 
These results suggest that prioritizing highly confident predictions improves user-level inference.

For both datasets, the relative advantage of confidence-based aggregation methods is notable for general texts. 
For GPT-4o on DDO general texts, the F1 score improves by 0.063 with the confidence-weighted method and by 0.066 with the maximum-confidence method compared with the text-level F1 score. 
On Reddit general texts, the improvements over the text-level score are even larger, reaching 0.166 and 0.193, respectively. 
GPT-4o on Reddit (Figure~\ref{fig:overview}g) achieves an F1 = 0.799 for general texts using the maximum-confidence method, representing a 0.193 improvement over the text-level F1 = 0.606. All pairwise differences among aggregation methods are statistically significant based on bootstrap-based paired t-tests ($p<0.01$).
These results indicate that LLMs can detect subtle signals of political alignment even in general-domain online discourse, achieving reliable inference across both datasets.

To contextualize these results, we further benchmarked LLMs against conventional supervised machine learning classifiers.
The maximum performance at the text level by the conventional models reaches 0.612 on the DDO dataset and 0.579 on the Reddit dataset (see Tables~\ref{tab:baseline_ddo} and \ref{tab:baseline_reddit} in Section~\ref{sec:benchmarking}).
Across both datasets, GPT-4o consistently outperforms most of the supervised baselines in both text- and user-level inference, with the gains over baselines being especially pronounced once predictions are aggregated at the user level.  
Incorporating confidence weighting further increased GPT-4o's user-level performance.
Llama-3.1-8B also generally matched or exceeded the best-performing traditional models. 
These findings highlight the strong inference capabilities of LLMs compared to conventional supervised approaches.

We also conducted a sensitivity analysis under stricter criteria for identifying general-topic texts, where potentially explicit political content was conservatively removed using trained political content classifiers (see Section~\ref{sec:sensitivity} in SI). 
Even after filtering, both LLMs maintained strong inference accuracy at the text and user levels.
These results suggest that our findings are robust.
Building on this foundation, we next examine the underlying factors that enable LLMs to extract political signals from general text and identify the topic categories and keywords that most strongly convey these signals.

\subsection{Variability in political inference across general topical categories}

Topics of ``Religion,'' ``Economics,'' ``Sports,'' and ``Arts'' differ in how strongly they encode political signals. Which domains are most informative for inferring political alignment? To answer this question, we evaluate inference performance across topical categories.

\begin{figure}
\centering
\includegraphics[width=\textwidth]{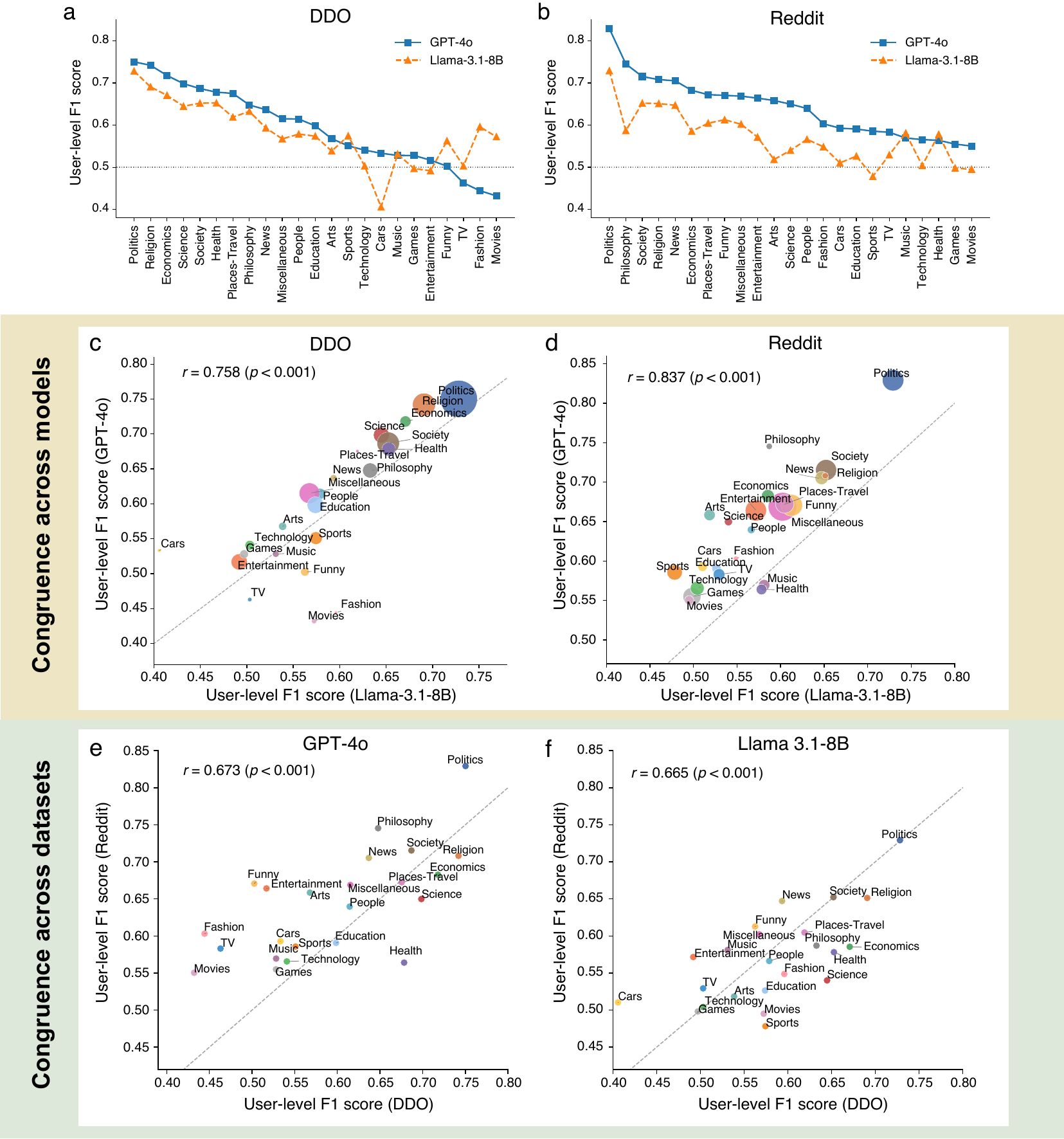}
\caption{\scriptsize
\textbf{Inference accuracy varies systematically across topical categories.} (a, b)~User-level F1 scores (using the maximum-confidence method) for political alignment inference across categories in (a)~DDO and (b)~Reddit using GPT-4o and Llama-3.1-8B. 
As a reference, the gray dotted lines in this and following figures indicate F1=0.5, the expected macro F1 score from a random classifier with balanced classes.
(c, d)~Correlations of user-level F1 scores by category between GPT-4o and Llama-3.1-8B for (c)~DDO and (d)~Reddit. 
Circle size is proportional to the number of observations in each category. 
(e, f)~Correlations of user-level F1 scores by category between DDO and Reddit for (e)~GPT-4o and (f)~Llama-3.1-8B. 
Prediction performance varies systematically across categories, independent of dataset or model.}
\label{fig:similarity_cat}
\end{figure}

Figure~\ref{fig:similarity_cat}a,b show that user-level F1 scores, computed with the maximum-confidence aggregation method, vary substantially across categories. As expected, ``Politics'' yields the highest inference performance in both datasets, with F1 scores of 0.750 for DDO and 0.829 for Reddit when inferred using GPT-4o. 
Beyond explicitly political discourse, several general categories such as ``Religion,'' ``Economics,'' ``Science,'' ``Society,'' and ``Health'' also enable strong inference. For instance, GPT-4o successfully infers a user's political alignment from their ``Religion'' texts in DDO with an F1 score of 0.742, and from ``Science'' and ``Health'' texts with scores of 0.698 and 0.687, respectively. In contrast, performance drops for categories such as ``Sports,'' ``Music,'' ``Fashion,'' and ``Movies''. 
These results reveal systematic differences in how strongly topics encode political signals. 
The same pattern emerges at the text-level F1 scores across datasets and LLMs (Figure~\ref{fig:category_text}), indicating that the observed performance variation across categories is driven by the informational content of the discourse itself, rather than differences in sample size or aggregation effects. 

We further examine whether the variation in inference performance across categories depends on the datasets or LLMs used for inference. 
Figures~\ref{fig:similarity_cat}c,d illustrate the alignment of category-level inference results between the two LLMs on the same datasets. 
For both DDO and Reddit, user-level F1 scores from GPT-4o and Llama-3.1-8B show strong positive correlations, with Pearson coefficients $r=0.758$ and $r=0.837$, respectively $(p<0.001)$, indicating that relative political inferability across categories is not driven by model-specific characteristics.

In addition, we compare category-level F1 scores for the same model across datasets (Figure~\ref{fig:similarity_cat}e,f). 
A model's performance for a given category on one platform correlates with its performance on the other, despite differences in platform norms, user composition, and discourse style. For instance, user-level F1 scores of GPT-4o across categories in DDO and Reddit yield a correlation of $r=0.673$ ($p<0.001$).
The stronger performance for entertainment-related categories on Reddit may reflect its more recent data and the growing politicization of lifestyle topics. 
Taken together with the cross-model alignment, this cross-dataset congruence suggests that variation in inference performance across categories reflects stable, discourse-level properties of categories themselves, rather than artifacts of specific models or platforms.


\subsection{Effects of semantic and social proximity on political inference}

\begin{figure}
\centering
\includegraphics[width=\textwidth]{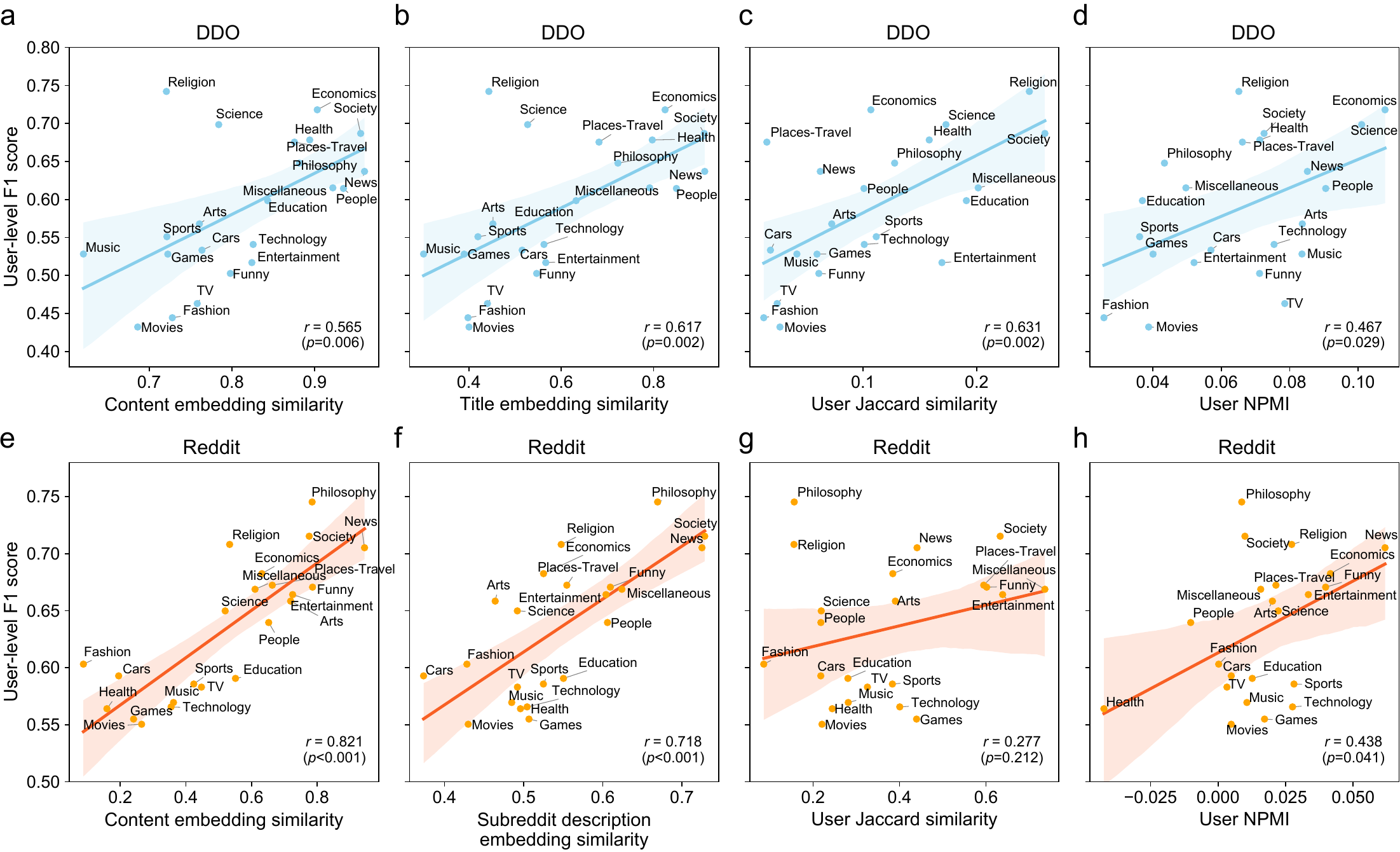}
\scriptsize
\caption{\scriptsize \textbf{Semantic similarity and user overlap with ``Politics'' predict higher inference performance across categories.} 
Panels~(a--d) show similarity measurements between general categories and the ``Politics'' category in the DDO dataset, based on (a)~content embedding similarity, (b)~debate title embedding similarity, (c)~Jaccard similarity of user participation between categories, and (d)~normalized pointwise mutual information (NPMI) of user participation. 
Content and title embedding similarities capture semantic similarity between categories, computed as the cosine similarity between category-level average embedding vectors obtained from a pre-trained Sentence-BERT model (\texttt{sentence-transformers/all-mpnet-base-v2}~\cite{song2020mpnet}). 
In contrast, Jaccard similarity and NPMI reflect user engagement overlap, quantifying the extent to which users participate in both categories.
Jaccard similarity ranges from 0 (no shared users) to 1 (identical user sets), and NPMI ranges from $-1$ (no shared users) to $+1$ (identical user sets), with zero indicating independence. 
Panels (e--h) show the same comparisons for the Reddit dataset. We use similarity in average subreddit description embeddings between categories (f) instead of title embeddings. 
GPT-4o is used for inference in all cases. 
Overall, categories with greater semantic similarity and user overlap with the ``Politics'' category tend to yield better inference in predicting political alignment.}
\label{fig:similarity_eng}
\end{figure}

The consistent differences in inference performance across topic categories, observed across both datasets and LLMs, raise an important question: Why are some categories more informative than others for inferring political alignment? 
We hypothesize that the answer lies in the associations between different categories and politics. 
To explore this hypothesis, we systematically examine two potential factors: the semantic similarity between a category's discourse and political content, and the extent of user participation overlap with political debates or subreddits.

Motivating the measurement of semantic similarity with politics, the text-as-data literature shows that ideological positioning can be detected in linguistic signals such as word choice and phrase usage, in corpora ranging from legislative speech~\cite{monroe2008fightin} to newspaper language~\cite{gentzkow2010drives}.
Beyond lexical characteristics, recent embedding-based approaches have demonstrated that political leaning can be inferred from semantic representations of text in diverse online domains. For instance, political bias can be detected by computing sentence-level embeddings of YouTube video titles~\cite{aldahoul2024polytc} or news articles~\cite{hong2023disentangling}.
These findings imply that categories whose discourse is semantically closer to political language should carry stronger signals for alignment inference.

On the social side, politically aligned users participate in multiple, partially overlapping communities and export their lexical and conversational norms across topics. 
Cross-topic authorship shows that stylistic ``fingerprints'' survive topic shifts \cite{sapkota2014cross} and social-media language reveals stable, identity-linked patterns at scale \cite{schwartz2013personality}, implying that political linguistic signals may spillover into other domains. 
Hence, categories with more politically engaged users should display stronger linguistic cues, and thus higher alignment-inference performance.

We first assess whether two forms of semantic similarity between a given category and the ``Politics'' category are associated with inference performance: the cosine similarity between (1)~average content embeddings and (2)~average debate title embeddings. For Reddit, instead of debate titles, we use subreddit descriptions. Embeddings are computed using the pre-trained Sentence-BERT model (\texttt{sentence-transformers/all-mpnet\-base-v2}~\cite{song2020mpnet}), which represents an input text as a 768-dimensional vector (See Methods). 
As shown in Figure~\ref{fig:similarity_eng}a,b, categories in DDO that are semantically closer to ``Politics'' tend to exhibit higher inference performance. Content embedding similarity correlates with user-level F1 scores calculated using GPT-4o and the maximum-confidence method (Pearson $r=0.565$, $p=0.006$).  Title embedding similarity shows an even stronger correlation ($r=0.617$, $p=0.002$). 
Similarly, in Reddit, we observe positive correlations between semantic similarity and inference performance (Figures~\ref{fig:similarity_eng}e,f). We obtain similar patterns using Llama-3.1-8B (Figure~\ref{fig:similarity_llama}). 
These results confirm that the semantic proximity of a topic to political discourse improves the performance of LLMs in inferring political alignment.
  
Next, we examine whether inference performance is associated with the degree of user overlap between a category and ``Politics.'' We compute Jaccard similarity and normalized pointwise mutual information (NPMI), both of which capture the extent to which the same users engage in both categories (see Methods). 
Note that in both DDO and Reddit, some users are never active in ``Politics'' (see Methods). 
This design allows us to test whether residual overlap between users engaged in political communities and those active in general-topic spaces provides stronger political alignment signals.

As shown in Figure~\ref{fig:similarity_eng}c,d for the DDO dataset, GPT-4o's performance is positively correlated with both user similarity metrics, Jaccard ($r=0.631$, $p=0.002$) and NPMI ($r=0.467$, $p=0.029$). Similar results are observed in the Reddit dataset (Figure~\ref{fig:similarity_eng}g,h), and for Llama-3.1-8B (Figure~\ref{fig:similarity_llama}). 
The weaker correlation for Reddit Jaccard similarity likely reflects the sampling structure of the Reddit dataset, where users were drawn from major political subreddits and thus do not represent the full population of participants in each topical category.
Together these results suggest that in categories with a higher prevalence of politically active users, political alignment is more readily inferred, likely because these participants carry distinctive linguistic features and conversational norms from political discourse. 

The semantic proximity of a category to political language and the political engagement of its users are not fully independent: politically active users can shape the semantic character of discourse, leading to positive correlations between semantic and social proximity measures (Figure~\ref{fig:similarity_corr_mat}). 
Both measures help us interpret why some categories are more informative than others for inferring political alignment.

\subsection{Lexical signals of partisanship in general discourse}

\begin{figure}
\centering
\includegraphics[width=\textwidth]{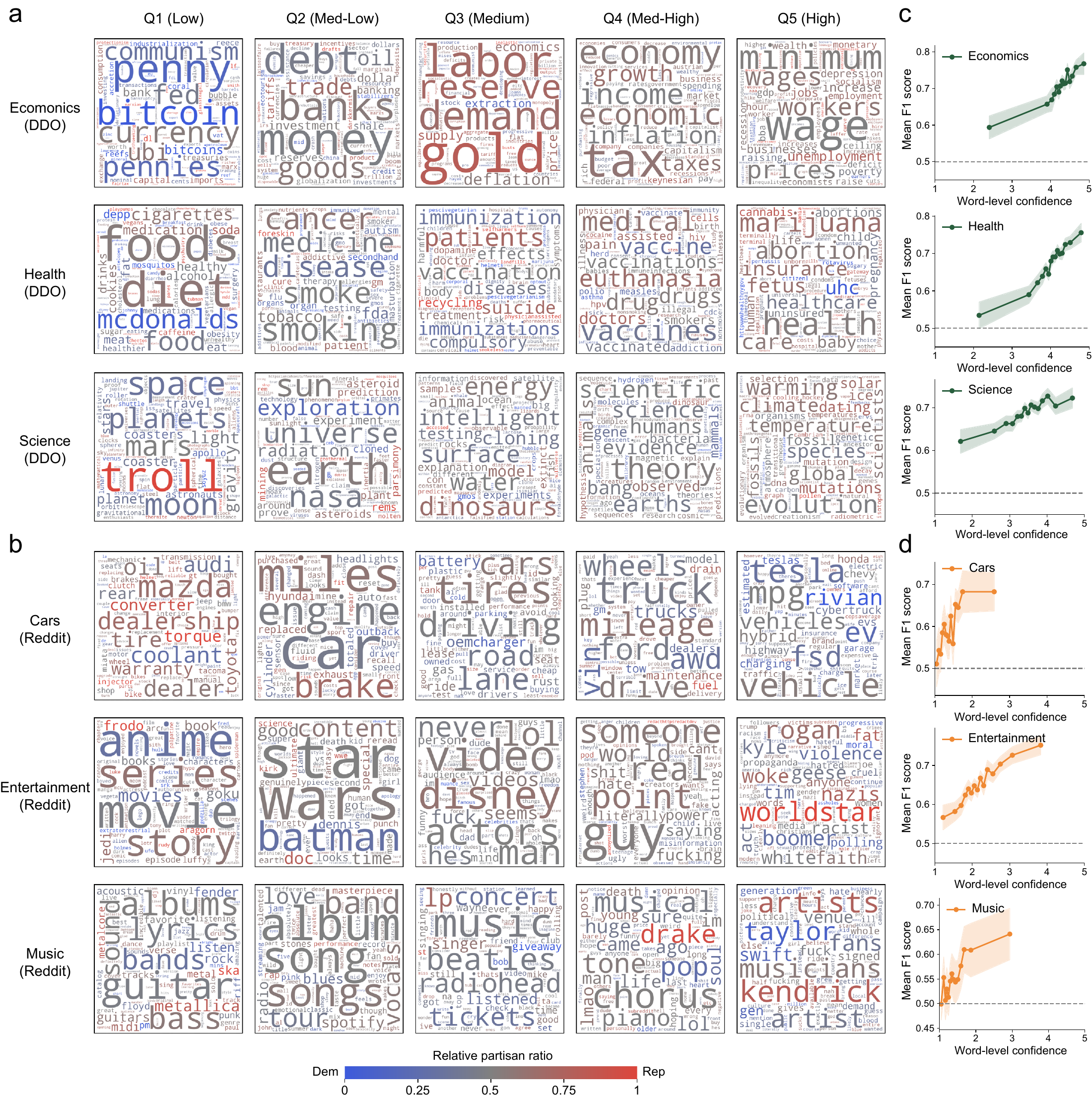}
\caption{\scriptsize
\textbf{Word-level confidence reveals politicization patterns in general-topic categories.} 
(a, b)~Six illustrative categories are shown: (a)~Economics, Health, and Science from the DDO dataset, and (b)~Cars, Entertainment, and Music from the Reddit dataset. Within each category, word clouds are arranged left to right by quintiles of word-level confidence, from lowest (Q1) to highest (Q5). Word-level confidence is defined as the mean confidence, averaged over all texts containing a given word, in inferring political alignment using GPT-4o. 
Higher confidence indicates stronger politicization signals, contributing more to LLM inference accuracy; even general-topic words in the Q5 clouds carry partisan signals. 
To visualize lexical distinctiveness, word size reflects the standardized log-odds $z$-score estimated using the Dirichlet prior method~\cite{monroe2008fightin}. 
Within each confidence quintile, we display up to the 100 words with the highest $z$-score values.
Word color indicates relative partisan ratio $f \in [0,1]$, where red ($f=1$) indicates exclusive usage by Republicans and blue ($f=0$) indicates exclusive usage by Democrats (see Methods).
(c, d)~Relationship between word-level confidence and mean F1 score for political alignment inference within each category. Each point represents one of 15 confidence quantiles, with shaded areas indicating 95\% confidence intervals. Across categories, higher-confidence words consistently yield more accurate inference.
These results demonstrate that LLMs effectively capture both implicit and explicit lexical cues relevant to political alignment.}
\label{fig:wordclouds}
\end{figure}

Although we have seen that inference performance in a given topical category is associated with the proximity of that category to politics, we wish to explore the effects of specific lexical cues within a category. 
Here we ask, which words enable LLMs to detect political signals in general discourse? Do general-topic categories contain words that, despite not being explicitly political, become politically charged and decisive for inference? If so, which words matter most, and how can we systematically characterize the strength of their partisan signal or their potential degree of politicization? To investigate this, we conduct a fine-grained analysis of word-level confidence signals across categories.

We begin with quantifying word-level confidence as the mean model-reported confidence for political classification, averaged across all texts within a category containing a given word (See Methods). A higher word-level confidence indicates that the LLM infers political alignment of texts in which the word appears with greater certainty. 
Figure~\ref{fig:wordclouds}a,b display six illustrative categories from DDO (Economics, Health, and Science) and Reddit (Cars, Entertainment, and Music), chosen to highlight words with varying levels of confidence (See Figures~\ref{fig:siwc_ddo1}--\ref{fig:siwc_reddit3} for results across all categories).  
Within each category, words are grouped into quintiles (Q1–Q5) by word-level confidence, from lowest to highest. 
However, because confidence distributions differ by category, a given quintile may represent different absolute ranges. 

Low-confidence quintiles tend to be populated by category terms that carry little partisan signal (e.g., ``penny'' (Economics), ``foods'' (Health), ``space'' (Science), and ``car'' (Cars) in Figure~\ref{fig:wordclouds}a,b). 
By contrast, high-confidence quintiles feature terms tied to ideological conflict or cultural polarization. 
Explicitly political or politicized words like ``tax'' (Economics), ``abortion'' (Health), and ``climate'' (Science) cluster at the upper end, together with culturally loaded expressions like ``Tesla'' (Cars), ``boomer'' (Entertainment), and ``Taylor Swift'' (Music), which often serve as cues of partisan leaning in online discourse. 

In the word clouds of Figure~\ref{fig:wordclouds}, color encodes the ratio of Democrats versus Republicans using a word. 
LLMs do not rely only on words with highly imbalanced usage frequencies, such as ``EV'' or ``Rivian'' in Cars (Reddit) and ``Taylor Swift'' in Music (Reddit), which are used disproportionately by Democrats. 
Rather, many of the most informative high-confidence words reflect a balanced usage across partisan groups, such as ``abortion'' in Health (DDO) or ``minimum wage'' in Economics (DDO). 
Thus, LLM predictions are not driven simply by lexical frequency differences between partisan groups. 
Rather, the models appear to leverage contextual and associative cues embedded in the contexts and associations through which words are deployed across partisan discourse.

Figure~\ref{fig:wordclouds}c,d quantify these patterns by plotting word-level mean F1 scores across 15 confidence quantiles, providing a finer-grained view of the relationship between confidence and inference performance. Here, the F1 score for each word is computed from the model's political inference performance on all texts in the category where the word appears. 
Across all categories, higher-confidence words consistently yield stronger partisan signals and thus more accurate classification (see Figure~\ref{fig:conf-f1-multi} for results across all categories in both datasets). 

Taken together, these results demonstrate that word-level confidence provides a systematic lens into how LLMs detect political signals in general discourse and helps identify the words that carry strong political signals. Higher-confidence words yield more accurate predictions of political alignment. Critically, these signals emerge not only from explicitly political vocabulary but also from subtle lexical cues embedded in everyday discourse. 

\section{Discussion}

This study investigates the ability of LLMs to infer the political alignment of individuals from general online discourse. 
By analyzing text from two distinct platforms, we show that advanced LLMs extract politically relevant signals from language use even in ostensibly nonpolitical settings. 
This approach advances prior work on digital trace data and computational inference~\cite{kosinski2013private, youyou2015computer} by demonstrating that political leaning is detectable outside of overtly political contexts without relying on prior user data for training. 
We make further methodological contributions by comparing text-level and user-level inference, incorporating confidence-based aggregation, and decoding the lexical cues that contribute to the predictions of LLMs. 

LLM-based inference has the potential to advance social and political science. 
In particular, our results point to promising opportunities for applying this pipeline to study the politicization of culture and everyday discourse. 
For instance, the ability of LLMs to capture partisan associations with terms such as ``eminem'' or ``swifties'' suggests a way to trace how cultural or seemingly nonpolitical topics become politically charged over time. 
Such applications could provide valuable insight into the evolving intersections of politics, culture, and communication in the digital age.

Our findings also point to significant risks for individual privacy. In particular, inference of political alignment can be exploited by enabling political micro-targeting and tailored messaging~\cite{borgesius2018online, simchon2023online}. 
Even privacy-aware users may become vulnerable by publicly sharing nonpolitical content that can be used to infer their political inclinations. 
Without updated privacy regulations, such AI applications risk amplifying online manipulation \cite{varol2017early,pacheco2021coordinated, torres2022deletions, menczer2024conversation} and further undermining trust in democratic processes.

Beyond these implications, several limitations should be considered when interpreting our results. 
First, our analyses rely on data from specific online platforms (DDO and Reddit), each with its own culture of participation and strong self-selection dynamics. 
For the DDO subset analyzed in this study, only users who voluntarily disclose their political identity are included, while on Reddit, users are labeled based on activity in highly partisan subreddits.
Due to these sampling constraints, our study primarily captures users who are more willing than average to reveal or signal their alignment, and therefore may not generalize to broader populations.

Second, the validity of our inferences is shaped by the properties of the LLMs we employ. 
Predictions may reflect inductive biases embedded in the training corpora.
Because the details of these corpora are undisclosed, it is difficult to disentangle whether the signals in our evaluation stem from patterns related or unrelated to user texts, that were internalized during training. This leaves our results open to criticism about hidden biases in the training data~\cite{gallegos2024bias}. 
Our findings may also depend on the particular models selected. 
We focused on GPT-4o and Llama-3.1-8B because of their complementary strengths in accuracy and reproducibility, yet different LLM architectures could yield different outcomes.
Future work should therefore assess the robustness of political inference across a broader range of models and training regimes.

Finally, our analyses are limited to English-language and U.S. contexts. 
This cultural and linguistic focus constrains the generality of our findings, since partisanship and the textual cues associated with it may differ substantially across political systems and societies. 
Future research could investigate whether the patterns we observe extend to other cultural settings, languages, and forms of political expression.

In sum, our findings highlight a dual implication. On one hand, our results demonstrate the utility of LLMs for detecting subtle social signals in everyday communication, opening new avenues for research on the associations between language use and personal attributes. On the other hand, they expose how easily private political orientation can be inferred from general content, raising serious concerns about individual privacy and the potential misuse of LLMs for micro-targeting. Acknowledging these limitations and risks is essential for both advancing scholarship and informing public debate on the societal implications of AI.

\section{Data and methods}
\label{sec:methods}


\subsection{Debate.org dataset}

We utilized the publicly available Debate.org (DDO) dataset \cite{durmus2018exploring, durmus2019corpus}, which was downloaded on March 18, 2023, from \url{https://esdurmus.github.io/ddo.html}. The original dataset consists of 78,376 debates by 42,906 debaters from October 15, 2007, to September 19, 2018.

Each debate in the dataset involves two participants: one arguing in favor of the proposition (PRO) and the other against it (CON). Debates are organized into multiple rounds, where participants present their arguments. For this study, all arguments made by a single user in a debate were aggregated into a single text, which was then provided as input for LLMs to infer the user's political alignment. Additionally, the DDO dataset includes metadata, including a user's self-reported political party alignment. Out of 3,511 debaters who disclosed their political party preference, 1,735 identified as Democratic, and 1,776 as Republican. For our study, we only used observations of these users. In total, we analyzed 22,265 arguments across 18,602 debates.

The dataset also includes a user-assigned category label for each debate, covering 23 categories such as ``Politics,'' ``Religion,'' ``Economics,'' ``TV,'' ``Game,'' ``Fashion,'' and others (Figure~\ref{fig:category_dist}). 
We considered debates labeled under the ``Politics'' category as political debates, all others as general discussions.

\subsection{Reddit dataset}


We used the Reddit API (\url{https://www.reddit.com/dev/api/}) to collect user comments, creating a dataset comparable to the DDO dataset in format. 
Each subreddit features user-submitted posts and corresponding comments, which can be upvoted or downvoted by other users. 
These votes generate a score for each comment, calculated as the number of upvotes minus downvotes. 

Unlike DDO, Reddit does not provide explicit metadata on a user's self-reported political alignment. To infer this information, we focused on two major American partisan subreddits: r/Conservative, a right-leaning community with about 1.1 million subscribers, and r/democrats, a left-leaning community with about 469,000 members as of September, 2024.
We retrieved the most popular 1,000 monthly submissions from each of these two subreddits as of September 14, 2024 and collected all comments from these submissions. 
This yielded 34,177 comments by 5,734 users in r/Conservative and 82,551 comments by 18,736 users in r/democrats. 
We excluded users whose comments received negative scores on average, ensuring that the selected users were supported by the partisan community, signaling political alignment. (We show in Figure~\ref{fig:commentscore_f1} that the LLM's inference is affected only marginally when using more stringent filters on a user's mean comment score.)
After this filtering, we retained 5,085 users in r/Conservative and 18,015 users in r/democrats.   
We randomly sampled 1,000 users from each list and assigned political proxy labels based on subreddit alignment: ``Republican'' for r/Conservative and ``Democrat'' for r/democrats. 

Next, we collected up to 100 recent comments by each of the 2,000 labeled users across all subreddits. This process yielded 97,319 comments from the Republican users and 96,616 comments from the Democrat users. 
Comments that were too short ($<10$ tokens) or excessively long ($>1{,}000$ tokens) were filtered out using OpenAI's \texttt{tiktoken} tokenizer, which led to the exclusion of seven Republican users and one Democrat user. We also excluded comments from the two source subreddits (r/Conservative and r/democrats), that were used to define user political alignments. 

Given that most of Reddit comments are extremely short (89\% contain five tokens or fewer), we aggregated each user's comments within a subreddit into a single text, which served as our unit of analysis. 
The final dataset includes 19,412 texts from 993 Republican users and 26,548 texts from 999 Democrat users, spanning 5,818 and 6,736 subreddits, respectively.


Since the Reddit dataset does not come with predefined topic categories, we classified each subreddit as political or general-topic by analyzing its description, which provides a brief introduction to its theme. 
We used GPT-4o (\texttt{gpt-4o-2024-08-06}) to classify all subreddits into the same 23 topic categories as in the DDO dataset (Figure~\ref{fig:category_dist}). 
Subreddits classified under the ``Politics'' category were treated as political content, consistent with the DDO dataset.  

\subsection{Political alignment inference}

Given a user text, we employed LLMs to infer whether the user's views aligned more closely with the Republican or Democratic Party. 
We tested two models: \texttt{gpt-4o-2024-08-06} (GPT-4o~\cite{openai_gpt4o_2024}) from OpenAI and \texttt{Llama-3.1-8B-Instruct} (Llama-3.1-8B~\cite{grattafiori2024llama}) from Meta. 
GPT-4o, a proprietary model, is known for its high inference accuracy, while Llama-3.1-8B, an open-source model, is lightweight and can be run locally. These models are chosen to leverage their complementary strengths and to represent both open and closed-source LLM architectures.

Each model was instructed to respond in a structured JSON format with two fields: \texttt{party}, indicating whether the inferred alignment was ``Republican'' or ``Democratic,'' and \texttt{confidence}, an integer score from 1 to 5. 
GPT-4o reliably adhered to the required JSON format under zero-shot prompting, producing correctly formatted responses for 96.1\% of texts in DDO and 97.4\% in Reddit  (Table~\ref{tab:answer_ratio}). 
Llama-3.1-8B often failed to produce correctly formatted outputs using the same prompt. Thus, we provided examples illustrating the required output format. This prompt yielded consistent responses in 97.7\% and 98.3\% of texts in the two datasets, respectively (Table~\ref{tab:answer_ratio}). 
The prompts used for both models across the two datasets are illustrated in Figures~\ref{fig:prompt-gpt-ddo}--\ref{fig:prompt-llama-reddit}. 

\subsection{Category similarity scores}

We assessed semantic and social proximity between general categories and political discourse by computing four similarity measures with respect to the ``Politics'' category: (1)~content embedding similarity, (2)~title or description embedding similarity, (3)~user Jaccard similarity, and (4)~user normalized pointwise mutual information (NPMI). 

\textbf{Content embedding similarity:} We computed content embeddings for the full sets of debate texts in DDO and subreddit comments in Reddit, averaging them within each category to obtain category-level content embeddings. 
These embeddings were generated using the pre-trained Sentence-BERT model (\texttt{sentence-transformers/all-mpnet-base-v2}~\cite{song2020mpnet}), which encodes each text into a 768-dimensional vector.
Content similarity between a given category $A$ and the ``Politics'' category $B$ was defined as the cosine similarity between the mean embeddings:
\begin{equation}
    \text{sim}_{\text{content}}(A,B) = \frac{\mathbf{c}_A \cdot \mathbf{c}_B}{\|\mathbf{c}_A\|\|\mathbf{c}_B\|},
\end{equation}
where $\mathbf{c}_A$ and $\mathbf{c}_B$ are the mean content embedding vectors for categories $A$ and $B$, respectively.

\textbf{Title or description embedding similarity:} For DDO, we obtained the embedding vectors of all debate titles in each category, while for Reddit, we obtained the embedding vectors of all subreddit descriptions. We then calculated the mean embedding vector for each category. 
Title embedding similarity between a given category $A$ and the ``Politics'' category $B$ was then defined as the cosine similarity between their mean title or description embeddings:
\begin{equation}
    \text{sim}_{\text{title}}(A,B) = \frac{\mathbf{v}_A \cdot \mathbf{v}_B}{\|\mathbf{v}_A\|\|\mathbf{v}_B\|},
\end{equation}
where $\mathbf{v}_A$ and $\mathbf{v}_B$ are the mean title/description embedding vectors for $A$ and $B$, respectively.

\textbf{User Jaccard similarity:} For each category $A$, let $U_A$ denotes the set of users who participated in that category. The Jaccard similarity with the ``Politics'' category $B$ is defined as
\begin{equation}
    J(A,B) = \frac{|U_A \cap U_B|}{|U_A \cup U_B|}.
\end{equation}
Jaccard similarity ranges from 0 to 1, while cosine similarity ranges from $-1$ to 1.

\textbf{User NPMI:} We also computed the normalized pointwise mutual information between user participation in category $A$ and in ``Politics'' $B$. Let $P(A)$ and $P(B)$ denote the marginal probabilities of user participation in category $A$ and $B$, respectively, and $P(A,B)$ the joint probability. Pointwise mutual information (PMI) is defined as
\begin{equation}
    \text{PMI}(A,B) = \log \frac{P(A,B)}{P(A)P(B)}.
\end{equation}
We normalize PMI~\cite{bouma2009normalized} to obtain 
\begin{equation}
    \text{NPMI}(A,B) = -\frac{\text{PMI}(A,B)}{\log P(A, B)}.
\end{equation}
NPMI ranges from $-1$ to $+1$, where $-1$ indicates that the events never co-occur, $+1$ indicates perfect co-occurrence, and $0$ indicates independence.


\subsection{Word cloud generation}

We constructed word clouds for each category across the two datasets to examine the relationship between word-level confidence and relative prominence in general-topic discourse.
For each dataset, observations were preprocessed using standard tokenization, lower-casing, and stop-word removal. 
Words appearing fewer than a minimum frequency threshold of 10 were excluded.

Word-level confidence scores were obtained based on political alignment predictions from GPT-4o.
For each word $w$, we computed the mean model-reported confidence across all texts containing $w$. 
Within each category, words were then ranked by their average confidence and divided into quintiles (Q1–Q5, from lowest to highest). 

We computed a category-level standardized log-odds score using the Dirichlet prior method~\cite{monroe2008fightin}, which compares the focal category to the pooled background of all other texts, to quantify each word's distinctiveness within a category. 
Let $A$ denote a category, let $x_w^{A}$ be the count of word $w$ in $A$, and let $N^{A}=\sum_{w} x_w^{A}$ be the total token count in $A$. 
For the background, we define an informative Dirichlet prior in which each word receives a prior pseudo-count equal to its frequency in the background corpus, i.e.,\ $\alpha_{w}$ is the count of word $w$ in the background, and $\alpha_{0}=\sum_{w}\alpha_{w}$ is the total background token count.
The posterior log-odds for $w$ in category $A$ relative to this prior baseline (posterior minus prior log-odds) is
\begin{equation}
\label{eq:logodds}
\text{log-odds}_{w}
= \log\frac{x_w^{A}+\alpha_{w}}{\left(N^{A}-x_w^{A}\right)+\left(\alpha_{0}-\alpha_{w}\right)}
\;-\;
\log\frac{\alpha_{w}}{\alpha_{0}-\alpha_{w}} \, .
\end{equation}
An approximate standard error is
\begin{equation}
\label{eq:se}
\text{SE}\!\left(\text{log-odds}_{w}\right)
\;\approx\;
\sqrt{
\frac{1}{x_w^{A}+\alpha_{w}}
\;+\;
\frac{1}{\left(N^{A}-x_w^{A}\right)+\left(\alpha_{0}-\alpha_{w}\right)}
\;+\;
\frac{1}{\alpha_{w}}
\;+\;
\frac{1}{\alpha_{0}-\alpha_{w}}
} \; ,
\end{equation}
and the standardized score used for selecting words and their font sizes in the word clouds is
\begin{equation}
\label{eq:z}
z_{w}
=\frac{\text{log-odds}_{w}}{\text{SE}\!\left(\text{log-odds}_{w}\right)} \, .
\end{equation}
The value $z_w$ is computed once at the \emph{category} level (not per quintile).

Using these $z_w$, we then selected up to the 100 most distinctive words (highest 
$z_w$) within each confidence quintile for visualization. Word-level confidence determines grouping (quintile assignment), whereas $z_w$ determines which words appear (top-100 per quintile) and their relative prominence (word size).

Word colors in the word clouds represent the relative partisan ratio of each term, computed as a continuous frequency-based score $f \in [0,1]$. 
For each word $w$, we first calculate the within-party relative frequency of that word, defined as the proportion of its occurrences among all words authored by Republican users ($p^{\text{Rep}}_w$) and among all words authored by Democratic users ($p^{\text{Dem}}_w$). 
The partisan leaning score $f_w$ is then obtained as:
\[
f_w = \frac{p^{\text{Rep}}_w}{p^{\text{Rep}}_w + p^{\text{Dem}}_w},
\]
where 
\[
p^{\text{Rep}}_w = \frac{\text{count}^{\text{Rep}}_w}{\sum_{w'} \text{count}^{\text{Rep}}_{w'}}, 
\qquad
p^{\text{Dem}}_w = \frac{\text{count}^{\text{Dem}}_w}{\sum_{w'} \text{count}^{\text{Dem}}_{w'}}
\] and $\text{count}^{\text{Rep}}_w$ ($\text{count}^{\text{Dem}}_w$) denotes the total number of occurrences of word $w$ in texts authored by Republican (Democratic) users. Thus, $f_w=1$ (pure red) indicates that a word is exclusively concentrated in 
Republican-authored texts, whereas $f_w=0$ (pure blue) indicates exclusive concentration 
in Democratic-authored texts. Intermediate values correspond to more balanced usage 
across the two groups, with $f_w \approx 0.5$ appearing as gray.

\bibliographystyle{unsrt}
\bibliography{main}

\section*{Ethics}

This research investigates the ability of LLMs to infer political alignment from online discourse. 
It could therefore be used in applications like political micro-targeting, to deliver tailored messages to specific voter segments. 
Such methodologies present risks of misuse, especially if personal data is collected and used without informed consent or when inference is scaled across millions of users. 
We believe the risks of the present research are outweighed by its benefits---including raised public awareness of these very risks. 
The data we use in our analysis is publicly available from debate histories on Debate.org and Reddit user activity (comments on subreddits). We do not use or share any personal identifiers. 
This study was approved as exempt from review by the Indiana University IRB (Protocol~\#26023).

\section*{Data availability}

For replication purposes, our processed, anonymized dataset can be downloaded at \url{https://github.com/ByunghweeLee-IU/llm-political-inference}.

\section*{Code availability}

Replication code is available at \url{https://github.com/ByunghweeLee-IU/llm-political-inference}.

\section*{Acknowledgments}
Y.Y.A. is supported by the Global Humanities and Social Sciences Convergence Research Program through the National Research Foundation of Korea (NRF), funded by the Ministry of Education (2024S1A5C3A02042671). S.K. and F.M. are supported in part by the Swiss National Science Foundation (grant CRSII5\_209250) and the Knight Foundation. H.K. was supported by the Republic of Korea's MSIT (Ministry of Science and ICT), under the Global Research Support Program in the Digital Field Program (RS-2024-00425354) supervised by the IITP (Institute of Information and Communications Technology Planning \& Evaluation). 

\section*{Competing interests}

The authors declare that they have no competing interests.


\clearpage

\vspace*{1cm}
\begin{center}
    \LARGE \textbf{Supporting Information} \\
    \vspace{0.5cm}
    \small \textbf{SI Text, SI Figures, and SI Tables}
\end{center}
\vspace{1cm}



\setcounter{figure}{0}
\setcounter{section}{0}
\setcounter{table}{0}
\renewcommand{\thefigure}{S\arabic{figure}}
\renewcommand{\thetable}{S\arabic{table}}
\renewcommand{\thesection}{S\arabic{section}}


\section{Validation for political alignment of Reddit users}
\label{sec:reddit_valid}

We assumed that the political alignment of Reddit users who received a positive mean score in the two political subreddits (\texttt{r/Conservative} and \texttt{r/democrats}) is consistent with the partisanship of the respective subreddits (i.e., supporting either the Republican or the Democratic party). To validate this assumption, we conducted a human evaluation using a set of sampled comments. We used the same Reddit dataset that includes submissions and comments from each subreddit, prior to sampling 1,000 users. From this dataset, we randomly sampled 25 submissions from each subreddit and collected both the submission titles and one positively scored comment per submission. We then combined and shuffled them, resulting in 50 (title + comment) pairs.

Five of the authors independently classified the users who posted these comments into two partisan groups: ``R'' for Republican and ``D'' for Democrat. This procedure was designed to assess whether human annotators consistently perceived users with positively scored comments as supporters of the corresponding partisan group.

\begin{figure}
\centering
\includegraphics[width=0.9\textwidth]{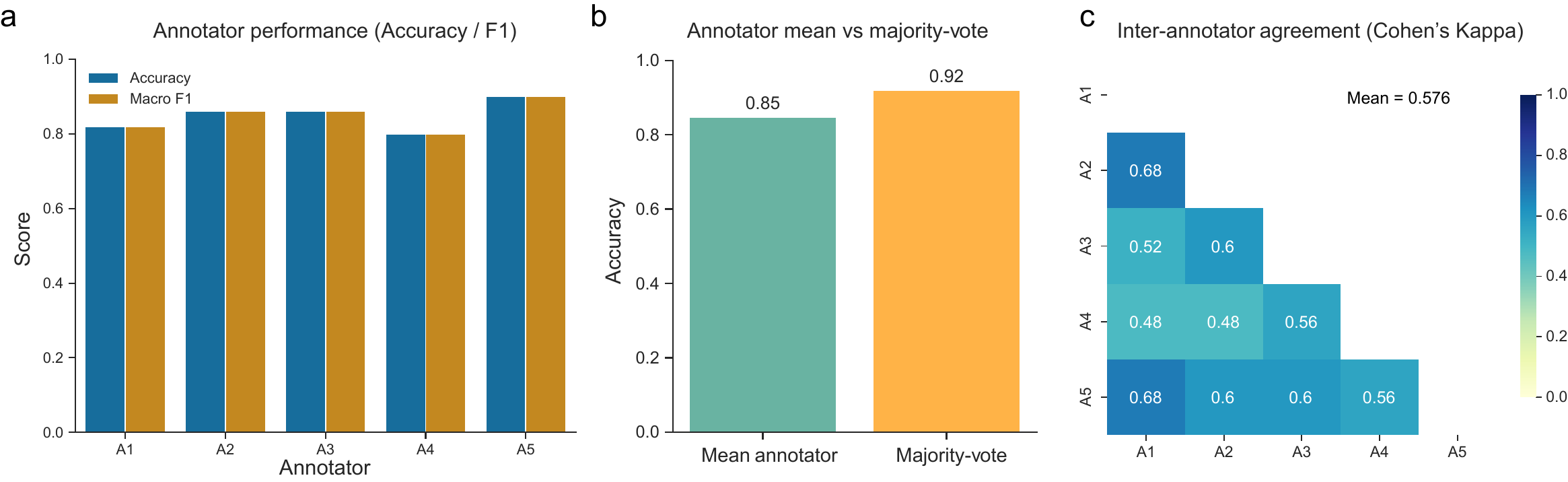}
\caption{(a) Performance of five annotators who independently inferred each user’s political alignment (Republican / Democrat) from 50 randomly sampled comments. Bars show individual accuracy and macro F1 scores, demonstrating consistently high classification performance across annotators.
(b) Comparison between the mean individual accuracy and the majority-vote accuracy against the true labels. The high majority-vote accuracy (0.92) and mean annotator accuracy (0.85) indicate that the heuristic labeling of a user's political alignment based on community activity is largely consistent with human judgment.
(c) Inter-annotator agreement measured by pairwise Cohen's $\kappa$ coefficients, showing a moderate level of consensus among annotators (mean $\kappa$ = 0.576). 
}
\label{fig:reddit_validation}
\end{figure}

The five annotators achieved a mean accuracy of 0.85 and a macro F1 score of 0.85, demonstrating consistently high classification performance (Figure~\ref{fig:reddit_validation}a).
When the labels are determined by majority vote among annotators, the accuracy further increases to 0.92 (Figure~\ref{fig:reddit_validation}b), indicating that the labeling of political alignment based on community activity is largely consistent with human judgment.
Figure~\ref{fig:reddit_validation}c reports the inter-annotator agreement measured by pairwise Cohen's~$\kappa$ coefficients, showing a moderate level of consensus among annotators (mean~$\kappa = 0.576$).
Fleiss'~$\kappa$, capturing the overall agreement among all annotators, was also~0.576, confirming that the annotators were reasonably consistent in their partisan inference.
Taken together, these results validate that users who receive positive scores within partisan communities are indeed likely to support the political orientation of those communities.

\section{Benchmarking against supervised machine learning baselines}
\label{sec:benchmarking}

\begin{table}
\centering
\scriptsize
\setlength{\tabcolsep}{10pt}
\caption{Political alignment inference results on the DDO dataset using traditional machine learning classifiers and LLMs.
We report average and standard deviation of macro F1 scores over 5-fold cross-validation, with 80\% of the data used for training and 20\% for testing in each fold. 
Text representations include TF-IDF and Sentence-BERT (S-BERT) embeddings, both applied across all traditional classifiers. 
For LLMs (GPT-4o and Llama-3.1-8B), predictions are made only on the test set from each fold, without access to training data. 
This approach ensures comparability with the traditional supervised models. 
Results are reported for three label types (Combined, Political, and General) at both the Text and User levels. 
For LLMs, user-level scores are aggregated from text-level predictions using different strategies (majority, confidence-weighted, and maximum-confidence), whereas traditional classifiers are evaluated directly at both levels.}
\begin{tabular}{llcccc}
\hline
\textbf{Model} & \textbf{Level} & \textbf{Combined} & \textbf{Political} & \textbf{General} \\
\hline
TF-IDF + NB (Multinomial) & Text & 0.579 $\pm$ 0.005 & 0.564 $\pm$ 0.015 & 0.583 $\pm$ 0.005 \\
& User & 0.546 $\pm$ 0.009 & 0.535 $\pm$ 0.009 & 0.553 $\pm$ 0.008 \\
S-BERT + NB (Gaussian) & Text & 0.544 $\pm$ 0.005 & 0.540 $\pm$ 0.011 & 0.542 $\pm$ 0.005 \\
& User & 0.540 $\pm$ 0.010 & 0.532 $\pm$ 0.006 & 0.532 $\pm$ 0.009 \\
TF-IDF + Logit & Text & 0.615 $\pm$ 0.007 & 0.623 $\pm$ 0.015 & 0.612 $\pm$ 0.008 \\
& User & 0.574 $\pm$ 0.014 & 0.583 $\pm$ 0.011 & 0.577 $\pm$ 0.015 \\
S-BERT + Logit & Text & 0.561 $\pm$ 0.004 & 0.562 $\pm$ 0.010 & 0.559 $\pm$ 0.003 \\
& User & 0.538 $\pm$ 0.006 & 0.537 $\pm$ 0.006 & 0.541 $\pm$ 0.010 \\
TF-IDF + SVM & Text & 0.608 $\pm$ 0.007 & 0.619 $\pm$ 0.015 & 0.603 $\pm$ 0.009 \\
& User & 0.573 $\pm$ 0.012 & 0.582 $\pm$ 0.017 & 0.565 $\pm$ 0.016 \\
S-BERT + SVM & Text & 0.568 $\pm$ 0.002 & 0.572 $\pm$ 0.010 & 0.566 $\pm$ 0.001 \\
& User & 0.545 $\pm$ 0.006 & 0.544 $\pm$ 0.007 & 0.540 $\pm$ 0.007 \\
TF-IDF + RF (n = 100) & Text & 0.597 $\pm$ 0.009 & 0.592 $\pm$ 0.014 & 0.599 $\pm$ 0.011 \\
& User & 0.545 $\pm$ 0.007 & 0.592 $\pm$ 0.014 & 0.599 $\pm$ 0.011 \\
S-BERT + RF (n = 100) & Text & 0.542 $\pm$ 0.007 & 0.541 $\pm$ 0.019 & 0.541 $\pm$ 0.005 \\
& User & 0.526 $\pm$ 0.004 & 0.530 $\pm$ 0.021 & 0.533 $\pm$ 0.005 \\
TF-IDF + XGB (n = 100) & Text & 0.593 $\pm$ 0.005 & 0.589 $\pm$ 0.012 & 0.594 $\pm$ 0.006 \\
& User & 0.554 $\pm$ 0.020 & 0.545 $\pm$ 0.024 & 0.560 $\pm$ 0.020 \\
S-BERT + XGB (n = 100) & Text & 0.548 $\pm$ 0.005 & 0.557 $\pm$ 0.011 & 0.544 $\pm$ 0.005 \\
& User & 0.540 $\pm$ 0.010 & 0.539 $\pm$ 0.009 & 0.543 $\pm$ 0.008 \\
\hline
GPT-4o & Text & 0.647 $\pm$ 0.004 & 0.711 $\pm$ 0.009 & 0.619 $\pm$ 0.005 \\
& User (majority) & 0.674 $\pm$ 0.006 & 0.727 $\pm$ 0.008 & 0.632 $\pm$ 0.006 \\
& User (confidence-weighted) & 0.692 $\pm$ 0.006 & 0.737 $\pm$ 0.010 & 0.650 $\pm$ 0.009 \\
& User (maximum-confidence) & 0.691 $\pm$ 0.008 & 0.737 $\pm$ 0.009 & 0.649 $\pm$ 0.009 \\
Llama-3.1-8B & Text & 0.619 $\pm$ 0.006 & 0.681 $\pm$ 0.004 & 0.595 $\pm$ 0.037 \\
& User (majority) & 0.646 $\pm$ 0.013 & 0.701 $\pm$ 0.007 & 0.612 $\pm$ 0.012 \\
& User (confidence-weighted) & 0.655 $\pm$ 0.016 & 0.710 $\pm$ 0.009 & 0.619 $\pm$ 0.017 \\
& User (maximum-confidence) & 0.654 $\pm$ 0.017 & 0.709 $\pm$ 0.006 & 0.617 $\pm$ 0.015 \\
\hline
\end{tabular}
\label{tab:baseline_ddo}
\end{table}

\begin{table}
\centering
\scriptsize
\setlength{\tabcolsep}{10pt}
\caption{Political alignment inference results on the Reddit dataset using traditional machine learning classifiers and LLMs.
All results are macro F1 scores averaged over 5-fold cross-validation, with 80\% of the data used for training and 20\% for testing in each fold. 
Text representations include TF-IDF and Sentence-BERT (S-BERT) embeddings, both applied across all traditional classifiers. 
For LLMs (GPT-4o and Llama-3.1-8B), predictions are made only on the test set from each fold, without access to training data. 
This approach ensures comparability with the traditional supervised models. 
Results are reported for three label types (Combined, Political, and General) at both the Text and User levels. 
For LLMs, user-level scores are aggregated from text-level predictions using different strategies (majority, confidence-weighted, and maximum-confidence), whereas traditional classifiers are evaluated directly at both levels.}
\begin{tabular}{llcccc}
\hline
\textbf{Model} & \textbf{Level} & \textbf{Combined} & \textbf{Political} & \textbf{General} \\
\hline
TF-IDF + NB (Multinomial) & Text & 0.513 $\pm$ 0.004 & 0.595 $\pm$ 0.020 & 0.502 $\pm$ 0.004 \\
& User & 0.465 $\pm$ 0.008 & 0.583 $\pm$ 0.022 & 0.459 $\pm$ 0.008 \\
S-BERT + NB (Gaussian) & Text & 0.585 $\pm$ 0.002 & 0.633 $\pm$ 0.005 & 0.579 $\pm$ 0.003 \\
 & User & 0.631 $\pm$ 0.008 & 0.640 $\pm$ 0.014 & 0.619 $\pm$ 0.007 \\
TF-IDF + Logit & Text & 0.564 $\pm$ 0.004 & 0.659 $\pm$ 0.007 & 0.553 $\pm$ 0.004 \\
& User & 0.572 $\pm$ 0.008 & 0.661 $\pm$ 0.007 & 0.554 $\pm$ 0.008 \\
S-BERT + Logit & Text & 0.577 $\pm$ 0.003 & 0.652 $\pm$ 0.010 & 0.568 $\pm$ 0.004 \\
& User & 0.586 $\pm$ 0.006 & 0.652 $\pm$ 0.011 & 0.570 $\pm$ 0.008 \\
TF-IDF + SVM & Text & 0.561 $\pm$ 0.003 & 0.637 $\pm$ 0.006 & 0.553 $\pm$ 0.004 \\
& User & 0.587 $\pm$ 0.006 & 0.634 $\pm$ 0.009 & 0.573 $\pm$ 0.007 \\
S-BERT + SVM & Text & 0.579 $\pm$ 0.004 & 0.653 $\pm$ 0.011 & 0.570 $\pm$ 0.004 \\
& User & 0.589 $\pm$ 0.010 & 0.647 $\pm$ 0.013 & 0.582 $\pm$ 0.002 \\
TF-IDF + RF (n = 100) & Text & 0.525 $\pm$ 0.006 & 0.610 $\pm$ 0.018 & 0.514 $\pm$ 0.005 \\
& User & 0.487 $\pm$ 0.015 & 0.598 $\pm$ 0.025 & 0.480 $\pm$ 0.015 \\
S-BERT + RF (n = 100) & Text & 0.534 $\pm$ 0.004 & 0.602 $\pm$ 0.008 & 0.526 $\pm$ 0.005 \\
& User & 0.500 $\pm$ 0.013 & 0.594 $\pm$ 0.016 & 0.493 $\pm$ 0.009 \\
TF-IDF + XGB (n = 100) & Text & 0.534 $\pm$ 0.002 & 0.619 $\pm$ 0.018 & 0.523 $\pm$ 0.004 \\
& User & 0.505 $\pm$ 0.008 & 0.618 $\pm$ 0.019 & 0.492 $\pm$ 0.010 \\
S-BERT + XGB (n = 100) & Text & 0.561 $\pm$ 0.002 & 0.621 $\pm$ 0.022 & 0.554 $\pm$ 0.005 \\
& User & 0.582 $\pm$ 0.008 & 0.620 $\pm$ 0.020 & 0.571 $\pm$ 0.009 \\
\hline
GPT-4o & Text & 0.624 $\pm$ 0.005 & 0.776 $\pm$ 0.005 & 0.606 $\pm$ 0.006 \\
& User (majority) & 0.676 $\pm$ 0.012 & 0.776 $\pm$ 0.006 & 0.642 $\pm$ 0.014 \\
& User (confidence-weighted) & 0.735 $\pm$ 0.007 & 0.783 $\pm$ 0.006 & 0.693 $\pm$ 0.006 \\
& User (maximum-confidence) & 0.734 $\pm$ 0.009 & 0.786 $\pm$ 0.006 & 0.689 $\pm$ 0.003 \\
Llama-3.1-8B & Text & 0.534 $\pm$ 0.002 & 0.670 $\pm$ 0.011 & 0.520 $\pm$ 0.003 \\
& User (majority) & 0.548 $\pm$ 0.008 & 0.675 $\pm$ 0.013 & 0.529 $\pm$ 0.006 \\
& User (confidence-weighted) & 0.612 $\pm$ 0.005 & 0.686 $\pm$ 0.008 & 0.589 $\pm$ 0.003 \\
& User (maximum-confidence) & 0.664 $\pm$ 0.008 & 0.690 $\pm$ 0.011 & 0.640 $\pm$ 0.006 \\
\hline
\end{tabular}
\label{tab:baseline_reddit}
\end{table}

Given the observed strength of LLMs in zero-shot settings, we further compared their accuracy with those of conventional supervised machine learning (ML) approaches. 
We benchmarked several classifiers against LLMs by evaluating political alignment inference using DDO and Reddit datasets.
We measured F1 scores averaged over 5-fold cross-validation, with 80\% of the data used for training and 20\% for testing in each fold.
For the ML models, we represented texts as vectors using TF-IDF and Sentence-BERT (S-BERT) embeddings, then trained several classifiers---Naive Bayes (NB), Logistic Regression (Logit), Support Vector Machine (SVM), Random Forest (RF), and XGBoost (XGB)---and evaluated their performance on the test sets (Tables~\ref{tab:baseline_ddo} and~\ref{tab:baseline_reddit}). 
For all ML models and LLMs, we measured inference performance at both the text and user levels. Since ML models do not generate confidence scores, we used the majority average method when aggregating predictions for user-level inference.

Table~\ref{tab:baseline_ddo} shows that on the DDO test set, GPT-4o outperforms supervised ML models in both text- and user-level inference across all text types (general, political, and combined) --- except for the TF-IDF + Logit model on general texts, where the difference is not significant. 
Llama-3.1-8B outperforms other supervised ML models in user-level inference. 
Tests on the Reddit dataset (Table~\ref{tab:baseline_reddit}) reveal a similar trend: GPT-4o consistently outperforms traditional supervised ML models and Llama-3.1-8B attains higher user-level scores than other ML models. 
These results highlight the strong inference capabilities of LLMs without task-specific training, particularly at the user level, where they consistently outperform supervised ML baselines.

\clearpage

\section{Robustness after controlling for user activity}
\label{sec:user-downsampling}

On DDO, hyperactive users, those who participated in many debates on the platform, might disproportionately influence the inference results, potentially leading to overfitting to active users during model training. (We note that in the Reddit dataset, users tend to contribute a relatively homogeneous number of texts, since we collected an equal number of comments per user.) To address this concern, we conducted an additional experiment on DDO in which users with more than 10 posts were limited to a random sample of 10 before splitting the dataset into train and test sets. We then repeated the inference task described in the above.

Table~\ref{tab:baseline_ddo_undersample} presents the inference results on the DDO dataset based on the user-undersampled data. The macro F1 scores showed a slight decrease at both the text and user levels, but overall performance remained similar compared to Table~\ref{tab:baseline_ddo}. For example, the TF-IDF + Logistic Regression model achieved a text-level F1 score of 0.577 and a user-level F1 score of 0.572. In contrast, the text-level performance of GPT-4o and Llama-3.1-8B slightly increased after undersampling hyperactive users, with F1 scores of 0.628 and 0.601, respectively. This improvement may reflect the removal of texts from disproportionately active participants who engage with relatively minor topics that carry limited political signal, as they tend to participate in a wider range of debates. Overall, these results indicate that the benchmarking performance of both supervised ML models and LLMs remains robust even after controlling for user activity.

\begin{table}
\centering
\scriptsize
\setlength{\tabcolsep}{10pt}
\caption{Political alignment inference results on the DDO dataset using traditional machine learning classifiers and LLMs, based on an undersampled version of the data.
To address concerns about hyperactive users disproportionately influencing the results and potentially leading to overfitting during model training, we conducted an additional experiment in which users with more than 10 posts were limited to a random sample of 10 observations.
All results are macro F1 scores averaged over 5-fold cross-validation, with 80\% of the data used for training and 20\% for testing in each fold. 
Text representations include TF-IDF and Sentence-BERT (S-BERT) embeddings, both applied across all traditional classifiers. 
For LLMs (GPT-4o and Llama-3.1-8B), predictions are made only on the test set from each fold, without access to training data. 
This approach ensures comparability with the traditional supervised models. 
Results are reported for three label types (Combined, Political, and General) at both the Text and User levels. 
For LLMs, user-level scores are aggregated from text-level predictions using different strategies (majority, confidence-weighted, and maximum-confidence), whereas traditional classifiers are evaluated directly at both levels.}
\begin{tabular}{llcccc}
\hline
\textbf{Model} & \textbf{Level} & \textbf{Combined} & \textbf{Political} & \textbf{General} \\
\hline
TF-IDF + NB (Multinomial) & Text & 0.556 $\pm$ 0.006 & 0.559 $\pm$ 0.007 & 0.554 $\pm$ 0.008 \\
& User & 0.560 $\pm$ 0.009 & 0.558 $\pm$ 0.012 & 0.557 $\pm$ 0.009 \\
S-BERT + NB (Gaussian) & Text & 0.544 $\pm$ 0.006 & 0.544 $\pm$ 0.006 & 0.541 $\pm$ 0.027 \\
& User & 0.545 $\pm$ 0.005 & 0.535 $\pm$ 0.023 & 0.542 $\pm$ 0.004 \\
TF-IDF + Logit & Text & 0.572 $\pm$ 0.005 & 0.574 $\pm$ 0.005 & 0.577 $\pm$ 0.008 \\
& User & 0.578 $\pm$ 0.004 & 0.575 $\pm$ 0.015 & 0.572 $\pm$ 0.010 \\
S-BERT + Logit & Text & 0.544 $\pm$ 0.009 & 0.533 $\pm$ 0.019 & 0.545 $\pm$ 0.012 \\
& User & 0.548 $\pm$ 0.008 & 0.526 $\pm$ 0.021 & 0.547 $\pm$ 0.014 \\
TF-IDF + SVM & Text & 0.564 $\pm$ 0.005 & 0.562 $\pm$ 0.011 & 0.564
$\pm$ 0.007 \\
& User & 0.567 $\pm$ 0.004& 0.562 $\pm$ 0.011 & 0.568 $\pm$ 0.009 \\
S-BERT + SVM & Text & 0.546 $\pm$ 0.007 & 0.541 $\pm$ 0.019 & 0.546 $\pm$ 0.012 \\
& User & 0.553 $\pm$ 0.006 & 0.541 $\pm$ 0.019 & 0.548 $\pm$ 0.010 \\
TF-IDF + RF (n = 100) & Text & 0.554 $\pm$ 0.011 & 0.553 $\pm$ 0.015 & 0.553 $\pm$ 0.011 \\
& User & 0.555 $\pm$ 0.012 & 0.553 $\pm$ 0.016 & 0.550 $\pm$ 0.013 \\
S-BERT + RF (n = 100) & Text & 0.535 $\pm$ 0.008 & 0.528 $\pm$ 0.018 & 0.536 $\pm$ 0.005 \\
& User & 0.539 $\pm$ 0.013 & 0.528 $\pm$ 0.017 & 0.537 $\pm$ 0.008 \\
TF-IDF + XGB (n = 100) & Text & 0.554 $\pm$ 0.004 & 0.540 $\pm$ 0.023 & 0.558 $\pm$ 0.006 \\
& User & 0.555 $\pm$ 0.006 & 0.538 $\pm$ 0.027 & 0.558 $\pm$ 0.011 \\
S-BERT + XGB (n = 100) & Text & 0.534 $\pm$ 0.009 & 0.535 $\pm$ 0.019 & 0.533 $\pm$ 0.013 \\
& User & 0.538 $\pm$ 0.015 & 0.538 $\pm$ 0.015 & 0.535 $\pm$ 0.013 \\
\hline
GPT-4o & Text & 0.659 $\pm$ 0.009 & 0.730 $\pm$ 0.023 & 0.628 $\pm$ 0.007 \\
& User (majority) & 0.668 $\pm$ 0.012 & 0.730 $\pm$ 0.025 & 0.637 $\pm$ 0.012 \\
& User (confidence-weighted) & 0.678 $\pm$ 0.013 & 0.731 $\pm$ 0.026 & 0.645 $\pm$ 0.014 \\
& User (maximum-confidence) & 0.682 $\pm$ 0.015 & 0.733 $\pm$ 0.025 & 0.644 $\pm$ 0.017 \\
Llama-3.1-8B & Text & 0.630 $\pm$ 0.013 & 0.701 $\pm$ 0.022 & 0.601 $\pm$ 0.011 \\
& User (majority) & 0.636 $\pm$ 0.014 & 0.705 $\pm$ 0.021 & 0.605 $\pm$ 0.014 \\
& User (confidence-weighted) & 0.649 $\pm$ 0.020 & 0.708 $\pm$ 0.018 & 0.615 $\pm$ 0.015 \\
& User (maximum-confidence) & 0.652 $\pm$ 0.016 & 0.707 $\pm$ 0.017 & 0.616 $\pm$ 0.014 \\
\hline
\end{tabular}
\label{tab:baseline_ddo_undersample}
\end{table}

\clearpage

\section{Sensitivity analysis under stricter general-topic criteria}
\label{sec:sensitivity}

During the inference process, we treated texts labeled as ``Politics'' as political and all others as general, based on debate categories from DDO and inferred subreddit categories from Reddit. 
However, some texts under a general topic may still contain explicitly political keywords. 
To assess the robustness of LLM inference under stricter general-topic criteria, we therefore conducted an additional analysis in which potentially explicit political content was filtered from general-topic texts.

\begin{figure}[h]
\centering
\includegraphics[width=0.8\textwidth]{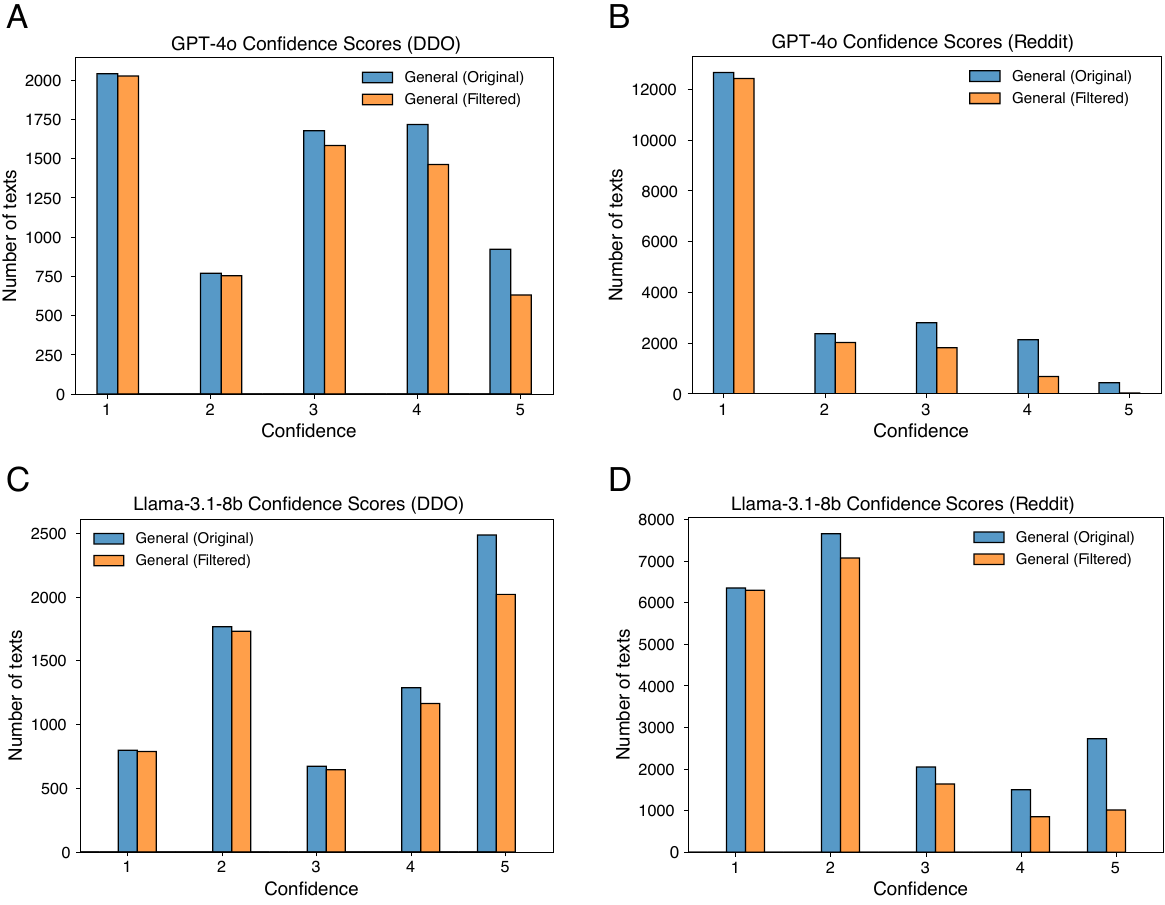}
\caption{Changes in confidence score distributions for general texts before and after filtering for sensitivity analysis, using GPT-4o and Llama-3.1-8B across two test sets (DDO and Reddit).
The plots show the distributions of confidence scores assigned by the two models to general texts, both before (blue) and after (orange) filtering out texts identified as explicitly political by a trained classifier. 
Filtering primarily removes high-confidence texts, particularly at the upper end of the confidence scale (4 or 5), suggesting that the classifier effectively detects political content embedded within general texts.}
\label{fig:conf-dist_sensitivity}
\end{figure}

To conservatively retain general texts, we trained political content classifiers using various ML models (Tables~\ref{tab:ddo_pol_classifier} and~\ref{tab:reddit_pol_classifier}). 
These classifiers were trained for binary classification of political texts from ``Politics'' and other categories. 
We then selected the best-performing classifier and applied it to the held-out half of the data to filter out potentially political texts from the general-topic set. 
The removed texts tend to have relatively higher LLM-reported confidence (Figure~\ref{fig:conf-dist_sensitivity}), suggesting that the classifier successfully identified texts that are similar to political contents.

After filtering out the texts which were classified as political texts in the test set by our classifier, we evaluated text- and user-level inference performance across all model–dataset pairs. 
As shown in Tables~\ref{tab:sensitivity_ddo} and~\ref{tab:sensitivity_reddit}, both LLMs maintained strong performance for both political and general texts. 
For example, GPT-4o achieves a text-level F1 score of 0.608 and a user-level F1 score of 0.670 using the maximum-confidence method on the DDO dataset (Table~\ref{tab:sensitivity_ddo}). On the Reddit dataset, Llama-3.1-8b shows a text-level F1 score of 0.528 and a user-level F1 score of 0.745 (Table~\ref{tab:sensitivity_reddit}). 
These findings show that LLMs can robustly infer political alignment even when potentially political content is conservatively removed from general-topic texts.

\begin{table}[h]
\centering
\small
\setlength{\tabcolsep}{10pt}
\caption{Political label classification results on the DDO dataset using traditional machine learning classifiers.
We used data from 50\% of the users for model development, reserving the remaining 50\% for a separate sensitivity analysis. 
Macro and weighted F1-scores with S.D. obtained from 5-fold cross-validation, with each fold using an 80/20 train-test split.
Text representations include TF-IDF and Sentence-BERT (S-BERT) embeddings, both applied across all traditional classifiers.
The best-performing classifier (in bold) was later used to assign political labels to text data in the held-out portion of the dataset.}
\begin{tabular}{lcc}
\hline
\textbf{Model} & \textbf{Macro F1 score} & \textbf{F1 Weighted} \\
\hline
TF-IDF + NB (Multinomial) & 0.782 $\pm$ 0.010 & 0.829 $\pm$ 0.008 \\
S-BERT + NB (Gaussian) & 0.784 $\pm$ 0.008 & 0.819 $\pm$ 0.008 \\
TF-IDF + Logit & 0.770 $\pm$ 0.007 & 0.826 $\pm$ 0.005 \\
S-BERT + Logit & 0.794 $\pm$ 0.005 & 0.841 $\pm$ 0.004\\
TF-IDF + SVM & 0.768 $\pm$ 0.010 & 0.821 $\pm$ 0.008 \\
\textbf{S-BERT + SVM} & \textbf{0.800 $\pm$ 0.006} & \textbf{0.845 $\pm$ 0.003} \\
TF-IDF + RF (n $=$ 100) & 0.773 $\pm$ 0.009 & 0.829 $\pm$ 0.006 \\
S-BERT + RF (n $=$ 100) & 0.752 $\pm$ 0.006 & 0.816 $\pm$ 0.005 \\
TF-IDF + XGB (n $=$ 100) & 0.767 $\pm$ 0.010 & 0.821 $\pm$ 0.009 \\
S-BERT + XGB (n $=$ 100) & 0.791 $\pm$ 0.007 & 0.839 $\pm$ 0.006 \\
\hline
\end{tabular}
\label{tab:ddo_pol_classifier}
\end{table}

\begin{table}[h]
\centering
\small
\setlength{\tabcolsep}{10pt}
\caption{Political label classification results on the Reddit dataset using traditional machine learning classifiers.
We used data from 50\% of the users for model development, reserving the remaining 50\% for a separate sensitivity analysis. 
Macro and weighted F1-scores with S.D. obtained from 5-fold cross-validation, with each fold using an 80/20 train-test split.
Text representations include TF-IDF and Sentence-BERT (S-BERT) embeddings, both applied across all traditional classifiers.
The best-performing classifier (in bold) was later used to assign political labels to text data in the held-out portion of the dataset.}
\begin{tabular}{lcc}
\hline
\textbf{Model} & \textbf{Macro F1 score} & \textbf{F1 Weighted} \\
\hline
TF-IDF + NB (Multinomial) & 0.507 $\pm$ 0.008 & 0.875 $\pm$ 0.010 \\
\textbf{S-BERT + NB (Gaussian)} & \textbf{0.673 $\pm$ 0.003} & \textbf{0.861 $\pm$ 0.005} \\
TF-IDF + Logit & 0.572 $\pm$ 0.014 & 0.887 $\pm$ 0.010 \\
S-BERT + Logit & 0.631 $\pm$ 0.008 & 0.897 $\pm$ 0.007 \\
TF-IDF + SVM & 0.612 $\pm$ 0.011 & 0.890 $\pm$ 0.007 \\
S-BERT + SVM & 0.626 $\pm$ 0.007 & 0.897 $\pm$ 0.007 \\
TF-IDF + RF (n $=$ 100) & 0.508 $\pm$ 0.008 & 0.875 $\pm$ 0.010 \\
S-BERT + RF (n $=$ 100) & 0.522 $\pm$ 0.013 & 0.878 $\pm$ 0.010 \\
TF-IDF + XGB (n $=$ 100) & 0.596 $\pm$ 0.005 & 0.889 $\pm$ 0.008 \\
S-BERT + XGB (n $=$ 100) & 0.617 $\pm$ 0.007 & 0.894 $\pm$ 0.006 \\
\hline
\end{tabular}
\label{tab:reddit_pol_classifier}
\end{table}

\begin{table}
\centering
\small
\setlength{\tabcolsep}{10pt}
\caption{F1 scores for general texts in the DDO dataset after filtering out potentially political content using the trained classifier.}
\begin{tabular}{llcccc}
\hline
\textbf{Model}&\textbf{Method} & \textbf{Combined} & \textbf{Political} & \textbf{General} \\
\hline
GPT-4o&Text-level           & 0.647 & 0.718 & 0.608 \\
&User (majority)        & 0.703 & 0.750 & 0.645 \\
&User (confidence-weighted)  & 0.735 & 0.769 & 0.673 \\
&User (maximum-confidence)       & 0.725 & 0.767 & 0.670 \\
\hline
Llama-3.1-8b&Text-level  & 0.637 & 0.690 & 0.615 \\
&User (majority)        & 0.686 & 0.717 & 0.645 \\
&User (confidence-weighted)  & 0.700 & 0.730 & 0.655 \\
&User (maximum-confidence)       & 0.710 & 0.737 & 0.657 \\
\hline
\end{tabular}
\label{tab:sensitivity_ddo}
\end{table}

\begin{table}
\centering
\small
\setlength{\tabcolsep}{10pt}
\caption{F1 scores for general texts in the Reddit dataset after filtering out potentially political content using the trained classifier.}
\begin{tabular}{llcccc}
\hline
\textbf{Model}&\textbf{Method} & \textbf{Combined} & \textbf{Political} & \textbf{General} \\
\hline
GPT-4o&Text level           & 0.626 & 0.768 & 0.578 \\
&User (majority)        & 0.731 & 0.842 & 0.613 \\
&User (confidence-weighted)  & 0.827 & 0.860 & 0.709 \\
&User (maximum-confidence)  & 0.850 & 0.859 & 0.722 \\
\hline
Llama-3.1-8b&Text-level  & 0.542 & 0.667 & 0.528 \\
&User (majority)        & 0.527 & 0.676 & 0.502 \\
&User (confidence-weighted)  & 0.671 & 0.726 & 0.646 \\
&User (maximum-confidence)       & 0.788 & 0.728 & 0.745 \\
\hline
\end{tabular}
\label{tab:sensitivity_reddit}
\end{table}

\clearpage
\section{Additional figures and tables}

\begin{figure}[h]
\centering
\includegraphics[width=\textwidth]{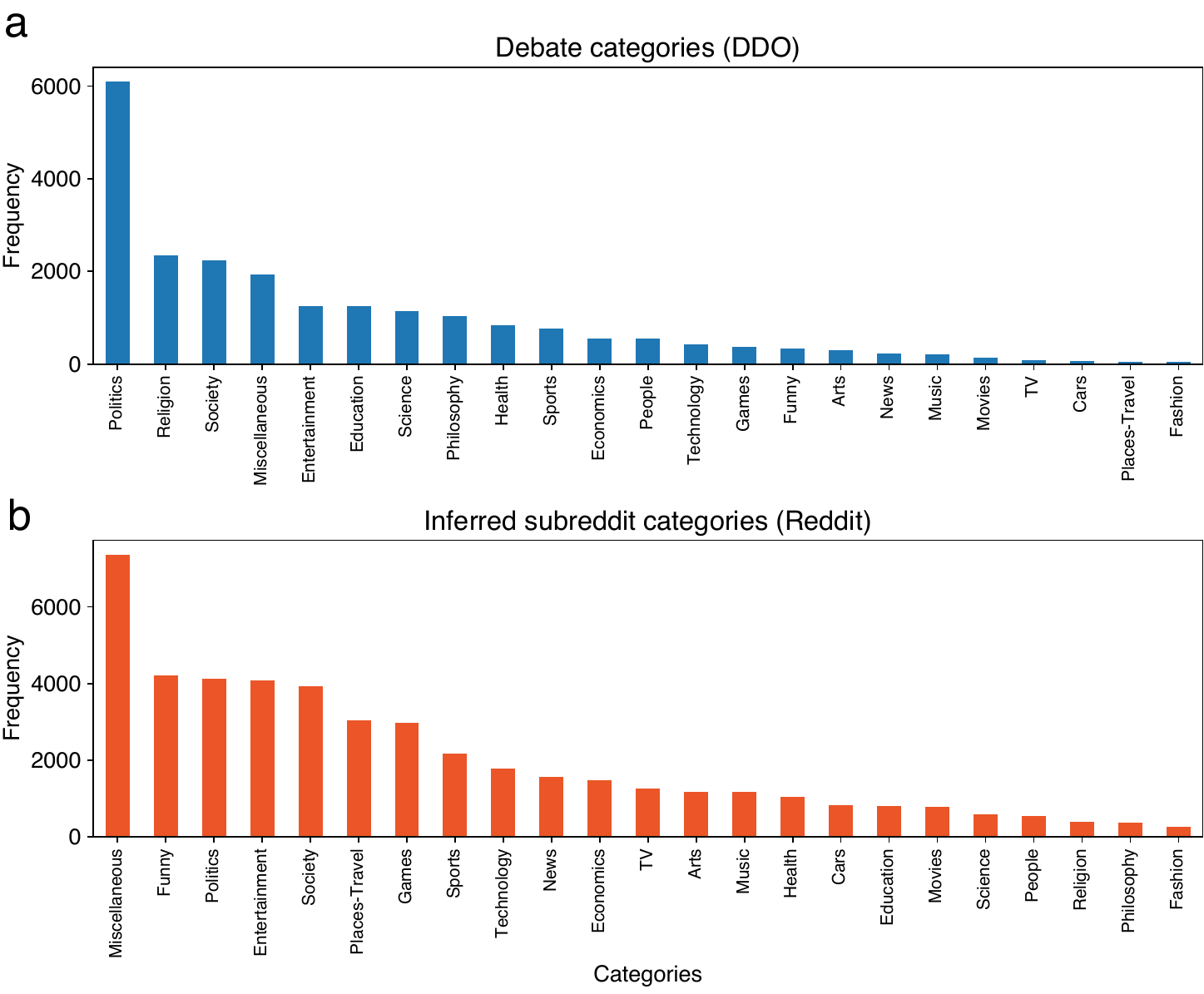}
\caption{Distribution of observations across 23 categories in the (a) DDO and (b) Reddit datasets. For Reddit, topic categories of communities (subreddits) were inferred from subreddit descriptions using GPT-4o and mapped to the same 23 categories as in DDO.}
\label{fig:category_dist}
\end{figure}

Figure~\ref{fig:category_dist} shows distributions of categories in the two datasets.

\clearpage
\begin{table}[p]
\centering
\small
\renewcommand{\arraystretch}{0.9}
\caption{\small Top 20 representative subreddits for each inferred category (23 categories in total), ranked by participation frequency in our dataset. Categories were inferred from subreddit descriptions using GPT-4o following the same category scheme as the DDO dataset.}
\label{tab:subreddit_examples}
\begin{tabular}{p{2.5cm}|p{14cm}}
\toprule
Category & Examples \\
\hline
\midrule
Arts & pics, books, comics, Cooking, DIY, suggestmeabook, woodworking, Infographics, interiordecorating, HomeDecorating, PhotoshopRequest, stories, DesignMyRoom, TattooDesigns, crochet, Tattoocoverups, tattoos, somethingimade, Broadway, Baking \\
\hline
Cars & CyberStuck, whatcarshouldIbuy, dashcams, motorcycles, MildlyBadDrivers, InfowarriorRides, electricvehicles, MechanicAdvice, cars, subaru, ToyotaTacoma, RealTesla, regularcarreviews, AskMechanics, BMW, thegrandtour, f150, Cartalk, AskAMechanic, classiccars \\
\hline
Economics & FluentInFinance, the\_everything\_bubble, Economics, wallstreetbets, economicCollapse, povertyfinance, jobs, unusual\_whales, personalfinance, economy, FirstTimeHomeBuyer, Money, inflation, REBubble, stocks, Accounting, StockMarket, Bitcoin, austrian\_economics, CryptoCurrency \\
\hline
Education & todayilearned, coolguides, explainlikeimfive, Teachers, USHistory, careerguidance, answers, Scams, careeradvice, education, AskHistory, howto, hiphop101, duolingo, highschool, WGU, cookingforbeginners, wikipedia, productivity, teaching \\
\hline
Entertainment & TikTokCringe, PublicFreakout, BoomersBeingFools, JoeRogan, nextfuckinglevel, OldSchoolCool, BeAmazed, Fauxmoi, StarWars, entertainment, popculturechat, CrazyFuckingVideos, videos, ThatsInsane, oddlysatisfying, CriticalDrinker, TwoHotTakes, NPR, Satisfyingasfuck, aww \\
\hline
Fashion & Watches, malefashionadvice, malegrooming, MakeupAddiction, mensfashion, handbags, whatthefrockk, weddingdress, fashion, Makeup, BusinessFashion, mensfashionadvice, EngagementRings, Weddingattireapproval, malehairadvice, beauty, Sneakers, OUTFITS, HairDye, fragrance \\
\hline
Funny & facepalm, clevercomebacks, AdviceAnimals, nottheonion, NewsOfTheStupid, funny, Jordan\_Peterson\_Memes, Wellthatsucks, memes, therewasanattempt, Unexpected, meirl, SipsTea, PeterExplainsTheJoke, RoastMe, Funnymemes, meme, MurderedByWords, ExplainTheJoke, maybemaybemaybe \\
\hline
Games & gaming, PS5, wow, Eldenring, Fallout, videogames, playstation, Helldivers, BaldursGate3, skyrim, DnD, Starfield, GeeksGamersCommunity, pcgaming, Steam, retrogaming, reddeadredemption, xbox, GTA6, fo4 \\
\hline
Health & ADHD, nursing, stopdrinking, 30PlusSkinCare, Semaglutide, DogAdvice, intermittentfasting, loseit, autism, LooksmaxingAdvice, sex, Residency, Drugs, Menopause, nutrition, Health, WegovyWeightLoss, crossfit, hygiene, alcoholicsanonymous \\
\hline
Miscellaneous & AskReddit, interestingasfuck, mildlyinfuriating, NoStupidQuestions, Damnthatsinteresting, mildlyinteresting, MadeMeSmile, MarkMyWords, conspiracy, MapPorn, cats, Bumperstickers, hypotheticalsituation, unpopularopinion, nostalgia, NonPoliticalTwitter, AMA, landscaping, MindBlowingThings, AllThatIsInteresting \\
\hline
Movies & movies, moviecritic, marvelstudios, Letterboxd, MovieSuggestions, shittymoviedetails, LV426, boxoffice, FIlm, flicks, underratedmovies, okbuddycinephile, comicbookmovies, lebowski, MovieLeaksAndRumors, deadpool, titanic, DC\_Cinematic, JamesBond, MarvelStudiosSpoilers \\
\hline
\bottomrule
\end{tabular}
\end{table}

\begin{table}[p]
\centering
\small
\renewcommand{\arraystretch}{0.9}
\caption{\small Top 20 representative subreddits for each inferred category (continued)}
\label{tab:subreddit_examples_cont}
\begin{tabular}{p{2.5cm} p{14cm}}\toprule
Category & Examples \\
\midrule
\hline
Music & Music, TaylorSwift, musicsuggestions, Guitar, hiphopheads, chappellroan, ClassicRock, KendrickLamar, MusicRecommendations, crappymusic, Bass, Eminem, beatles, LinkinPark, popheads, MetalForTheMasses, country, spotify, LetsTalkMusic, Foofighters \\
\hline
News & news, AnythingGoesNews, inthenews, worldnews, FOXNEWS, USNewsHub, UkraineWarVideoReport, OutOfTheLoop, UkrainianConflict, UpliftingNews, nytimes, southcarolina, Canada\_sub, PBS\_NewsHour, anime\_titties, Hunting, worldnewsvideo, offbeat, fednews, newzealand \\
\hline
People & AskMen, GenX, Asmongold, daddit, Destiny, AskOldPeopleAdvice, elonmusk, AirForce, USMC, SwiftlyNeutral, EntitledPeople, IAmA, GlowUps, Faces, musked, Serverlife, AmazonFC, GenXWomen, kardashians, FamousFaces \\
\hline
Philosophy & AmItheAsshole, changemyview, DecodingTheGurus, Anarcho\_Capitalism, JordanPeterson, whatif, Life, samharris, SeriousConversation, DeepThoughts, LiminalSpace, FutureWhatIf, Manipulation, ExplainBothSides, InsightfulQuestions, SimulationTheory, likeus, philosophy, TheNightFeeling, AskLibertarians \\
\hline
Places-Travel & texas, Ohio, Pennsylvania, minnesota, europe, florida, Michigan, Georgia, wisconsin, travel, ukraine, NatureIsFuckingLit, delta, California, TravelMaps, canada, missouri, Seattle, phoenix, massachusetts \\
\hline
Politics & Conservative, politics, PoliticalHumor, KamalaHarris, Presidents, LeopardsAteMyFace, walkaway, AntiTrumpAlliance, PoliticalCompassMemes, conservativeterrorism, Defeat\_Project\_2025, PoliticalDiscussion, trump, Firearms, ShitPoliticsSays, Fuckthealtright, libsofreddit, IntellectualDarkWeb, Law\_and\_Politics, Libertarian \\
\hline
Religion & atheism, religiousfruitcake, wholesomememes, Christianity, TrueChristian, Catholicism, FundieSnarkUncensored, Jewish, NoahGetTheBoat, exmormon, exchristian, Bible, Judaism, exmuslim, Christian, Christians, Reformed, OpenChristian, AskAChristian, mythology \\
\hline
Science & science, geography, Futurology, space, skeptic, aliens, whatsthisplant, arborists, animalid, CreationNtheUniverse, climate, telescopes, climatechange, environment, nuclear, climateskeptics, spiders, 23andme, whatsthisbug, insects \\
\hline
Society & WhitePeopleTwitter, AITAH, GenZ, Millennials, BlackPeopleTwitter, antiwork, FuckImOld, AmIOverreacting, law, insanepeoplefacebook, Xennials, millenials, TwoXChromosomes, relationship\_advice, AskOldPeople, recruitinghell, TrueUnpopularOpinion, Adulting, MURICA, MarchAgainstNazis \\
\hline
Sports & nfl, CFB, sports, nba, baseball, olympics, golf, fantasyfootball, formula1, ufc, SquaredCircle, hockey, NFLv2, mlb, soccer, cowboys, MMA, NBATalk, detroitlions, NCAAFBseries \\
\hline
TV & television, gameofthrones, HouseOfTheDragon, TheBoys, startrek, sitcoms, DunderMifflin, IASIP, TheSimpsons, seinfeld, southpark, BigBrother, rickandmorty, netflix, televisionsuggestions, americandad, 90DayFiance, thesopranos, TedLasso, 90dayfianceuncensored \\
\hline
Technology & technology, dataisbeautiful, pcmasterrace, ChatGPT, aviation, apple, iphone, 3Dprinting, sysadmin, gadgets, SteamDeck, GooglePixel, AppleWatch, technews, hvacadvice, buildapc, PleX, ITCareerQuestions, StableDiffusion, tmobile \\
\hline
\bottomrule
\end{tabular}
\end{table}

\begin{figure}[h]
\centering
\includegraphics[width=0.8\textwidth]{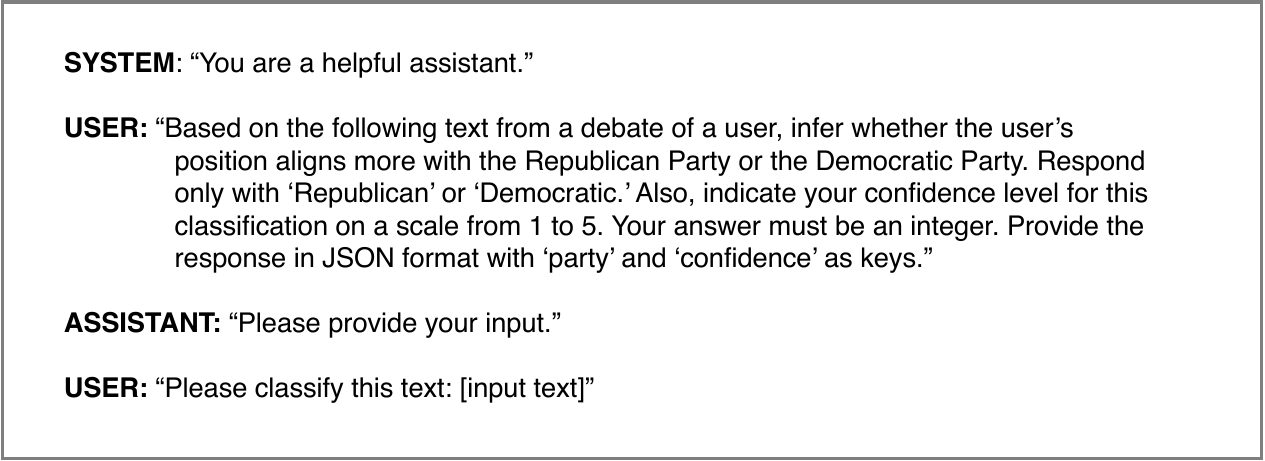}
\caption{The prompt used by GPT-4o for inferring a DDO user's political alignment.}
\label{fig:prompt-gpt-ddo}
\end{figure}

\begin{figure}[h]
\centering
\includegraphics[width=0.8\textwidth]{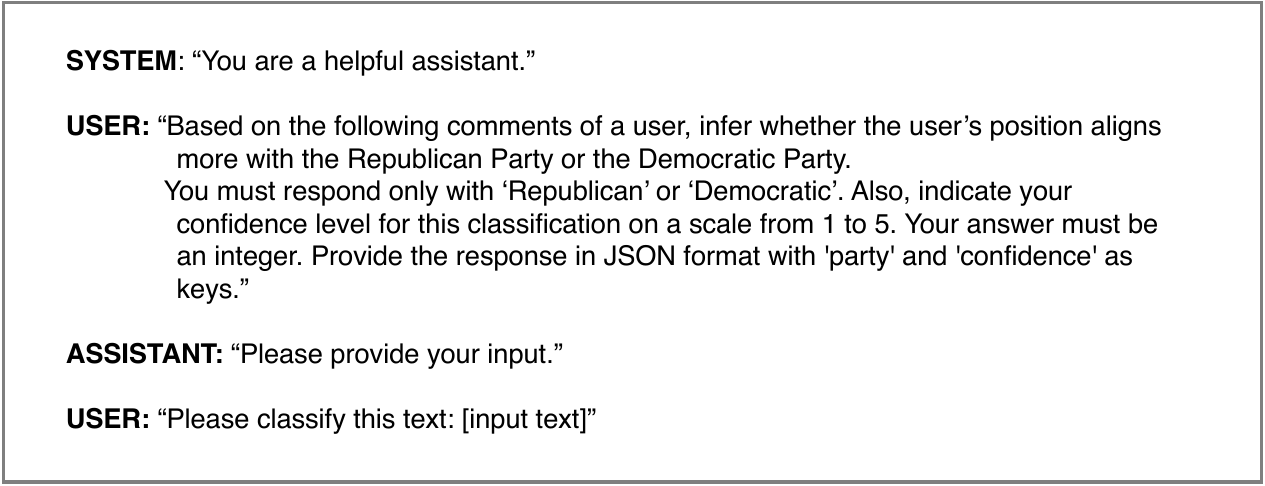}
\caption{The prompt used by GPT-4o for inferring a Reddit user's political alignment.}
\label{fig:prompt-gpt-reddit}
\end{figure}

\begin{figure}[h]
\centering
\includegraphics[width=0.8\textwidth]{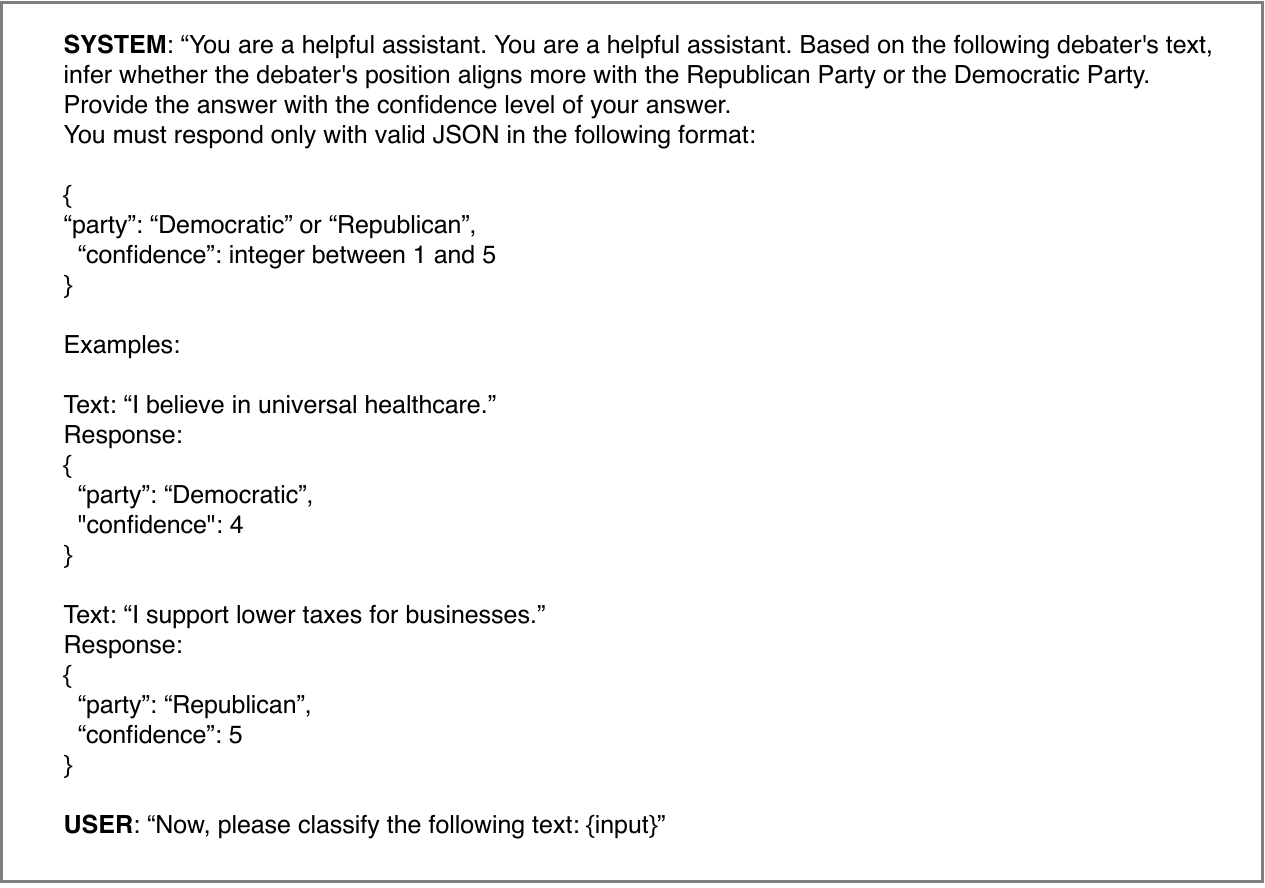}
\caption{Prompt used by Llama-3.1-8B for inferring a DDO user's political alignment.}
\label{fig:prompt-llama-ddo}
\end{figure}

\begin{figure}[h]
\centering
\includegraphics[width=0.8\textwidth]{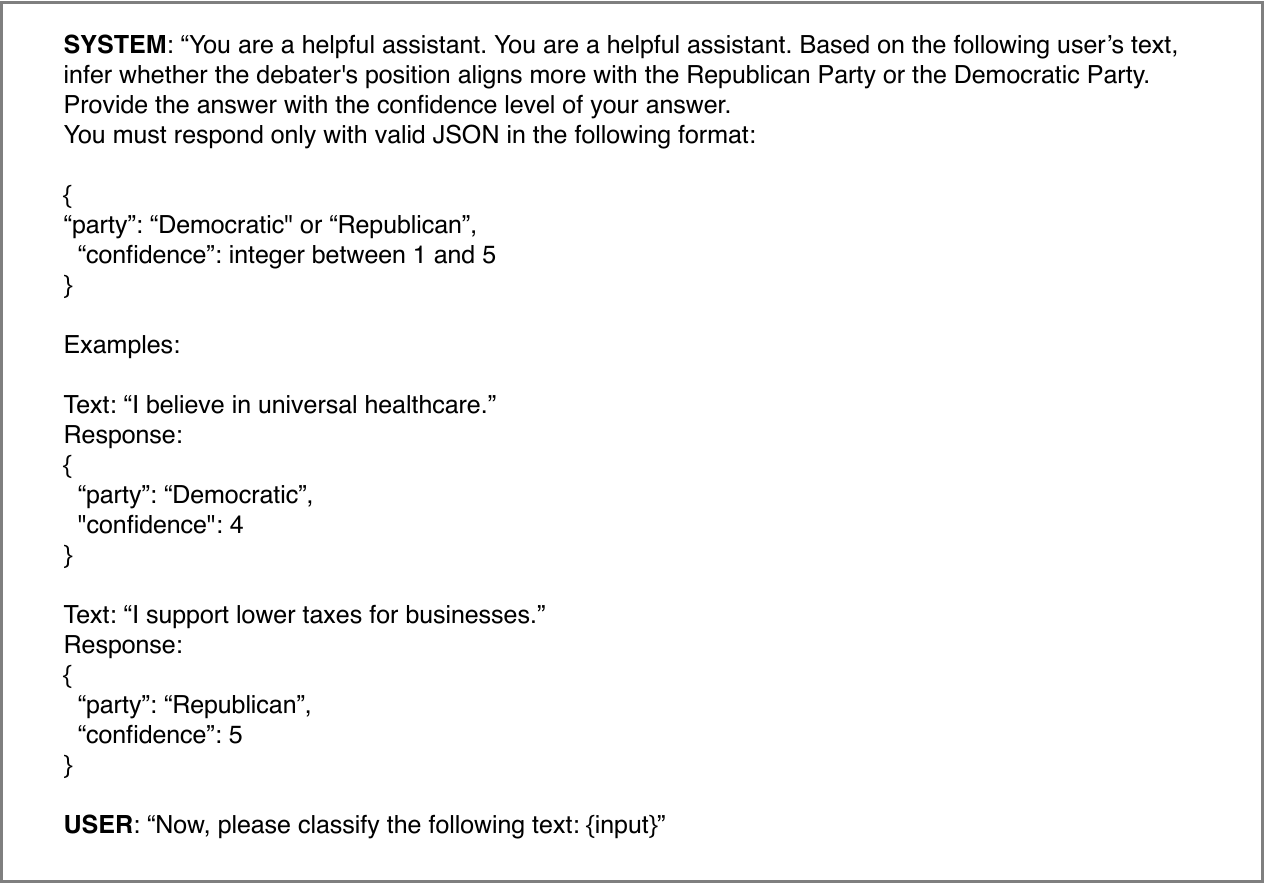}
\caption{Prompt used by Llama-3.1-8B for inferring a Reddit user's political alignment.}
\label{fig:prompt-llama-reddit}
\end{figure}

\clearpage
Figures~\ref{fig:prompt-gpt-ddo}--\ref{fig:prompt-llama-reddit} report on the prompts used for both models across the two datasets. 

\clearpage

\begin{figure}[h]
\centering
\includegraphics[width=0.8\textwidth]{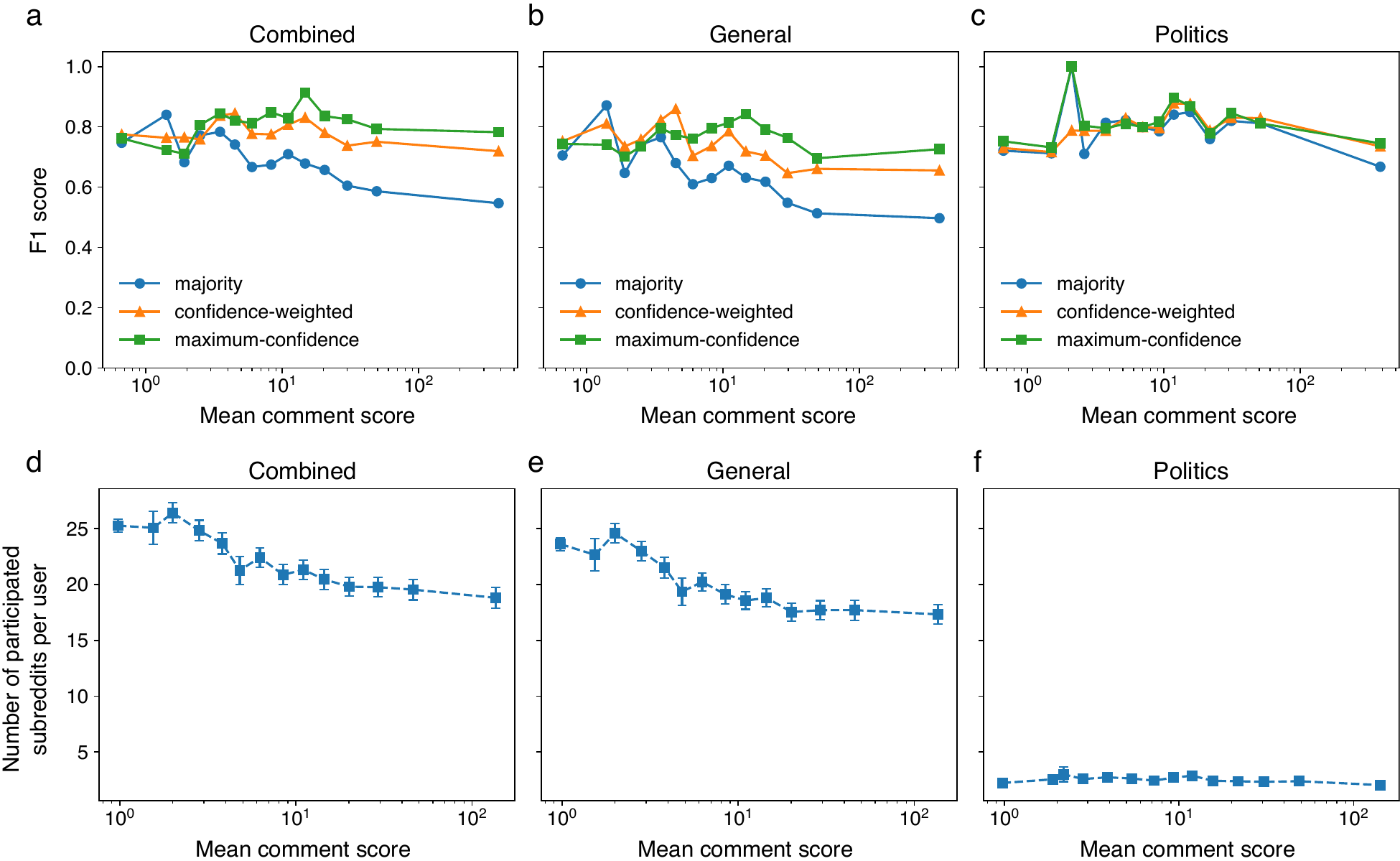}
\caption{\footnotesize User-level political alignment inference performance using GPT-4o across Reddit user groups with varying mean comment scores in the source communities (\texttt{r/Conservative} or \texttt{r/democrats}). 
To ensure that our preprocessing choice—retaining only users with positive mean comment scores—does not bias our results, we examine whether GPT-4o's user-level political alignment inference performance varies across the full range of mean comment scores. 
This analysis tests the robustness of our findings to alternative implicit threshold values for user inclusion.
During preprocessing, we retained only users who consistently received positive feedback, operationalized as having positive mean comment scores (upvotes minus downvotes) across their recent comments. 
We then grouped users by their mean comment scores to examine how inference performance of LLMs varies across user groups. 
(a--c) Macro F1 scores across mean comment score groups for three textual contexts: (a) Combined, (b) General, and (c) Politics. Each panel compares three user-level aggregation methods applied to GPT-4o predictions: (1) majority vote, (2) confidence-weighted average, and (3) maximum-confidence average. 
Across all three contexts, inference performance remains consistent across a user's mean comment scores.
The confidence-weighted and maximum-confidence aggregation methods show stable performance patterns throughout the entire score range, indicating that GPT-4o's user-level inference is robust to variations in mean comment scores.
Only the majority-vote method shows a decline for high–mean-score users in the Combined and General contexts, likely due to these users contributing fewer general-topic comments; however, this effect does not appear in the Politics context.
(d--f) Number of participated subreddits per user across the same mean comment score groups for (d) Combined, (e) General, and (f) Politics categories.
The decreasing trend in (d) and (e) demonstrates that users with high mean comment scores typically participate in a smaller, more concentrated set of subreddits, consistent with the performance drop observed for the majority-vote method in (a) and (b).
In the Politics category (f), the number of participated subreddits remains relatively stable across score groups, aligning with the more uniform performance patterns observed in (c). Symbols and error bars represent mean $\pm$ standard error.}
\label{fig:commentscore_f1}
\end{figure}

Figure~\ref{fig:commentscore_f1} shows the robustness of user-level political alignment inference across the full range of mean comment scores in the source communities. 

\clearpage

\begin{figure}[h]
\centering
\includegraphics[width=0.6\textwidth]{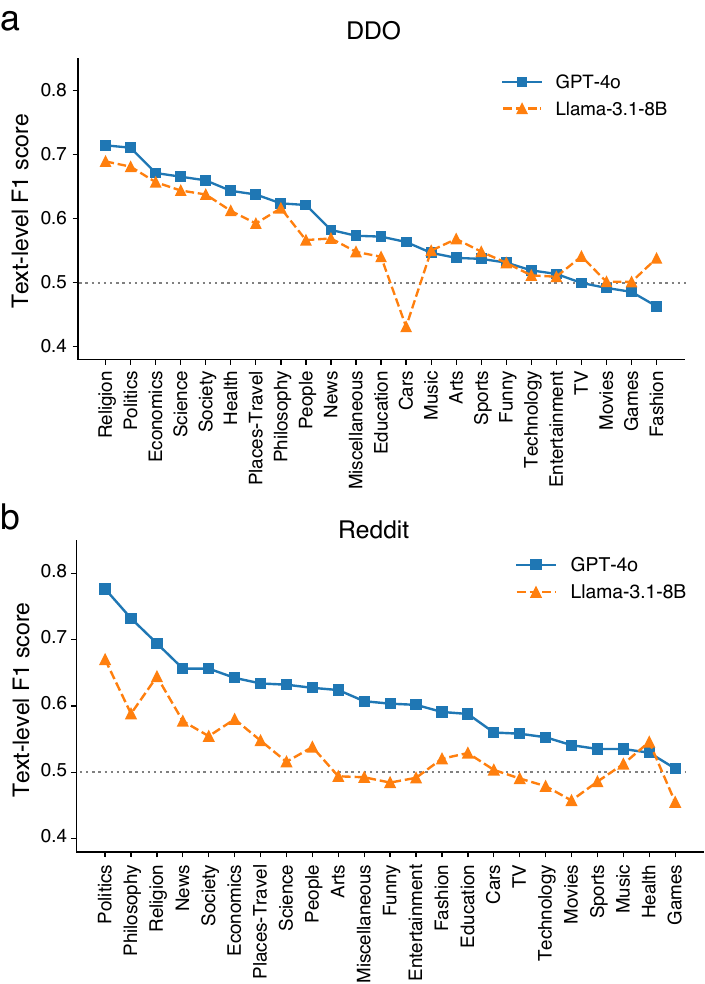}
\caption{(a) Text-level F1 scores for political alignment inference across categories in the DDO dataset for GPT-4o and Llama-3.1-8B. (b) Text-level F1 scores across categories in the Reddit dataset for GPT-4o and Llama-3.1-8B. The gray dotted lines indicate an F1 score of 0.5, representing the expected macro-F1 of a random classifier.
}
\label{fig:category_text}
\end{figure}

Figure~\ref{fig:category_text} plots political alignment inference performance across categories. 

\clearpage

\begin{figure}[h]
\centering
\includegraphics[width=\textwidth]{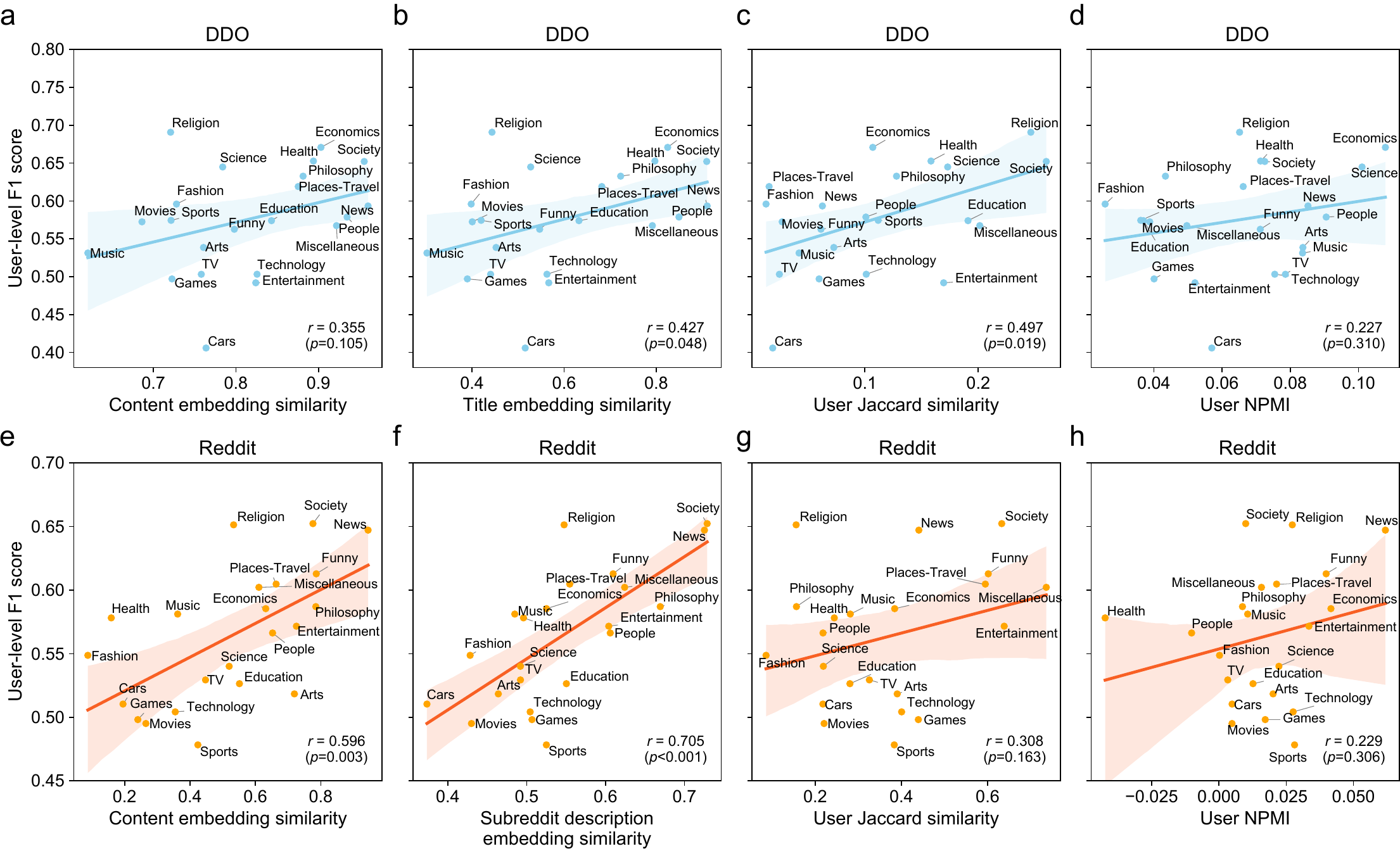}
\caption{
Effect of semantic similarity and user overlap with politics on inference performance across categories, based on user-level F1 scores calculated using the maximum-confidence method with Llama-3.1-8B.
(a--d) Similarity measurements between general categories and the Politics category in the DDO dataset, based on (a) content embedding similarity, (b) debate title embedding similarity, (c) Jaccard similarity of user participation between categories (User Jaccard similarity), and (d) normalized pointwise mutual information of user participation (User NPMI). (e--h) Same comparisons for the Reddit dataset. For Reddit, we use similarity in average subreddit descriptions embeddings between categories instead of title embeddings.
}
\label{fig:similarity_llama}
\end{figure}

Figure~\ref{fig:similarity_llama} shows how political alignment inference is affected by the semantic proximity of topics to political discourse using Llama-3.1-8B.

\clearpage

\begin{figure}[h]
\centering
\includegraphics[width=\textwidth]{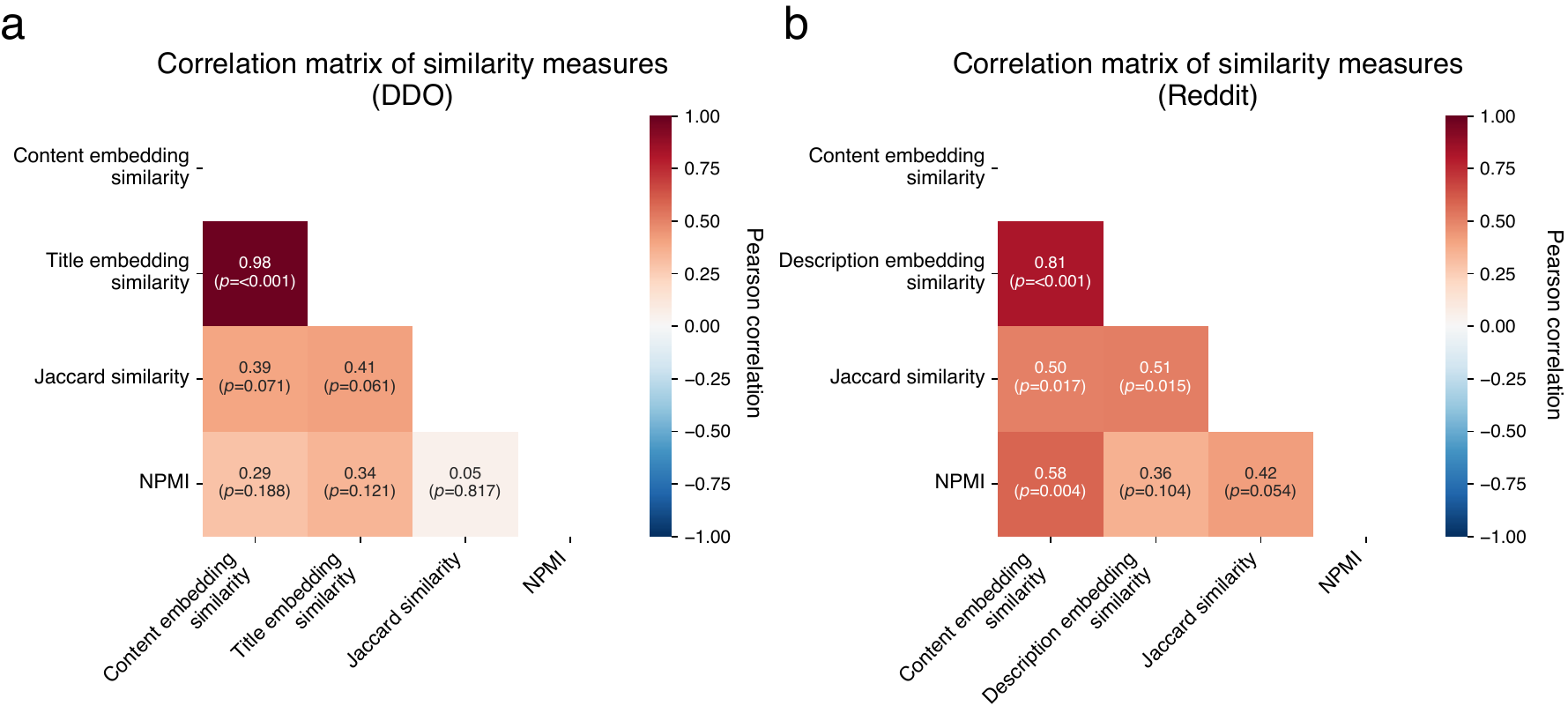}
\caption{ 
Correlation matrices between the four similarity measures (content embedding similarity, title or subreddit description embedding similarity, user Jaccard similarity, and user NPMI) for (a) the DDO dataset and (b) the Reddit dataset. Each similarity measure comprises 22 values corresponding to correlations between each category and the ``Politics'' category. Pearson correlation coefficients between pairs of similarity measures are shown, with the associated $p$-values in parentheses.
}
\label{fig:similarity_corr_mat}
\end{figure}

Figure~\ref{fig:similarity_corr_mat} reports correlations among different measures of similarity between categories and ``Politics.'' 

\clearpage

\begin{figure}[h]
\centering
\includegraphics[width=0.75\textwidth]{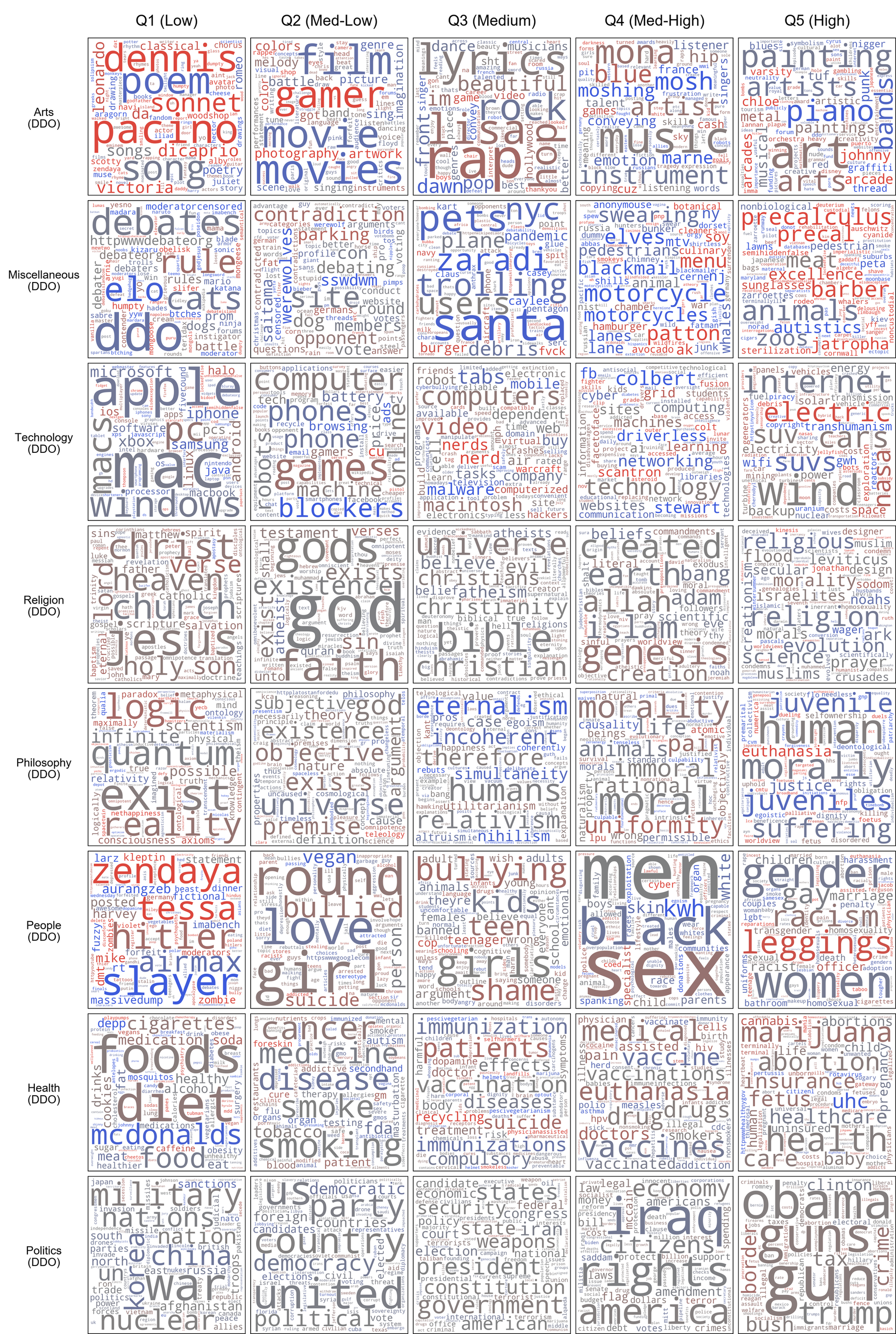}
\caption{DDO word clouds (Arts, Miscellaneous, Technology, Religion, Philosophy, People, Health, Politics)}
\label{fig:siwc_ddo1}
\end{figure}

\begin{figure}[h]
\centering
\includegraphics[width=0.8\textwidth]{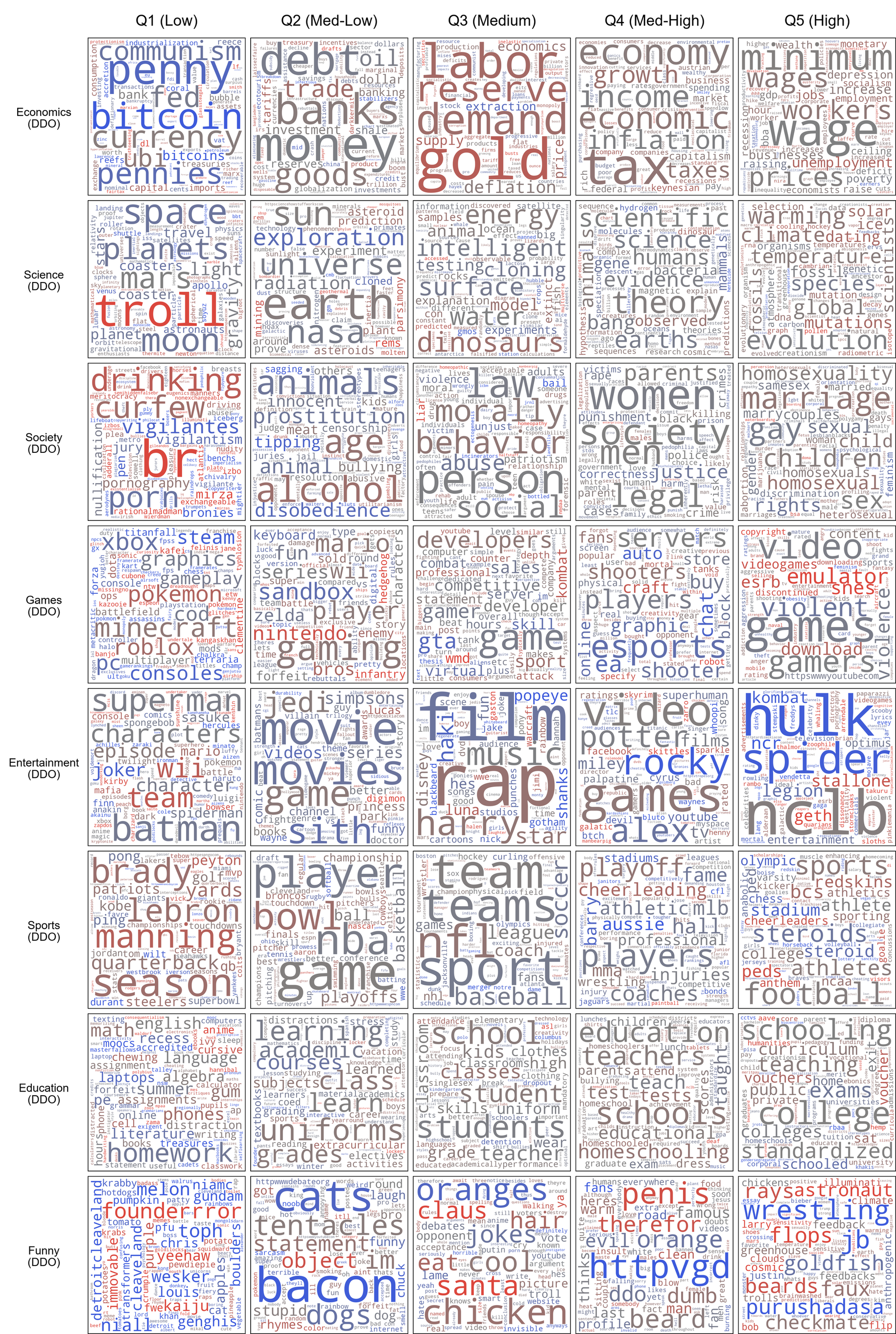}
\caption{DDO word clouds (Economics, Science, Society, Games, Entertainment, Sports, Education, Funny)}
\label{fig:siwc_ddo2}
\end{figure}

\begin{figure}[h]
\centering
\includegraphics[width=0.8\textwidth]{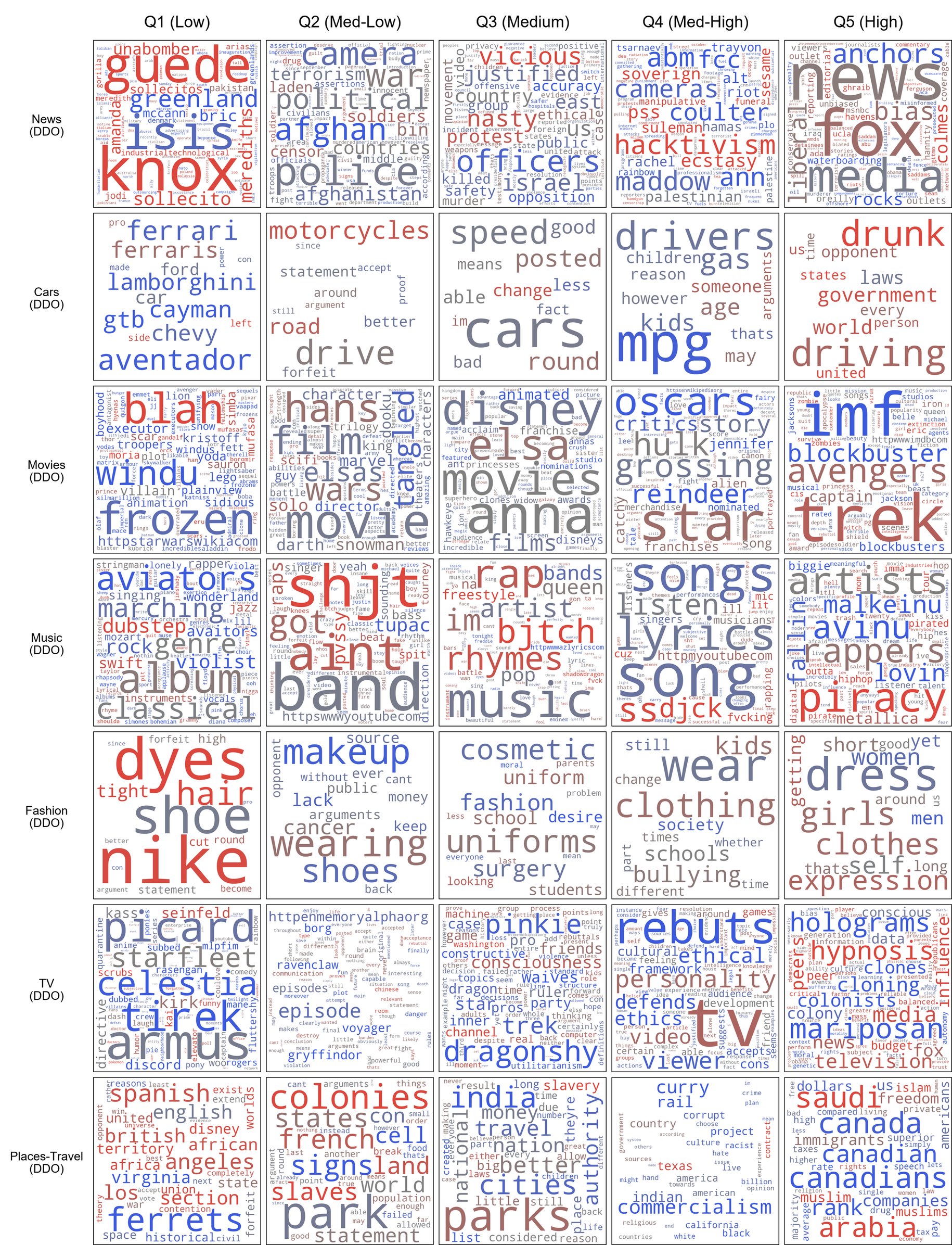}
\caption{DDO word clouds (News, Cars, Movies, Music, Fashion, TV, Places-Travel)}
\label{fig:siwc_ddo3}
\end{figure}

\begin{figure}[h]
\centering
\includegraphics[width=0.8\textwidth]{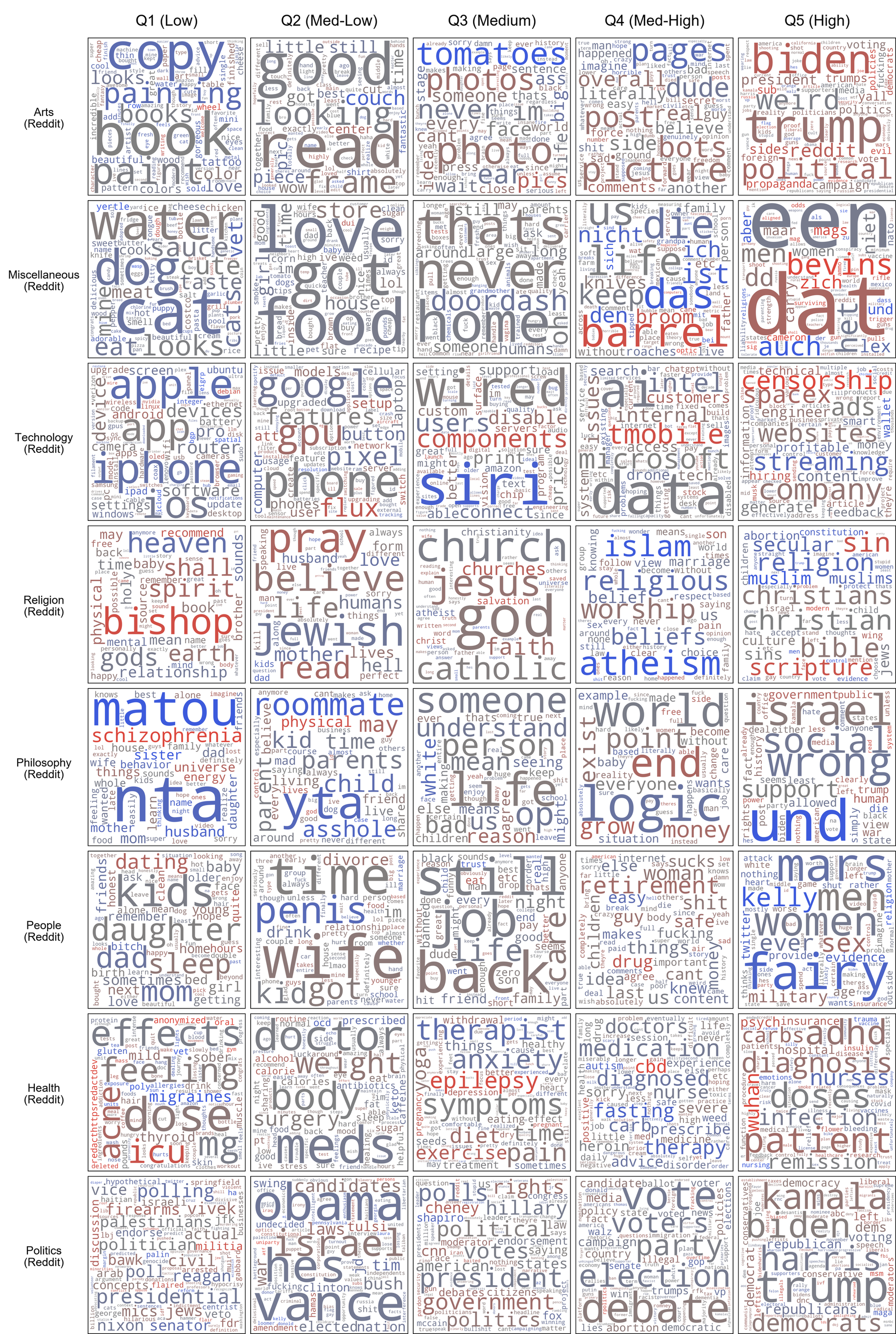}
\caption{Reddit word clouds (Arts, Miscellaneous, Technology, Religion, Philosophy, People, Health, Politics)}
\label{fig:siwc_reddit1}
\end{figure}

\begin{figure}[h]
\centering
\includegraphics[width=0.8\textwidth]{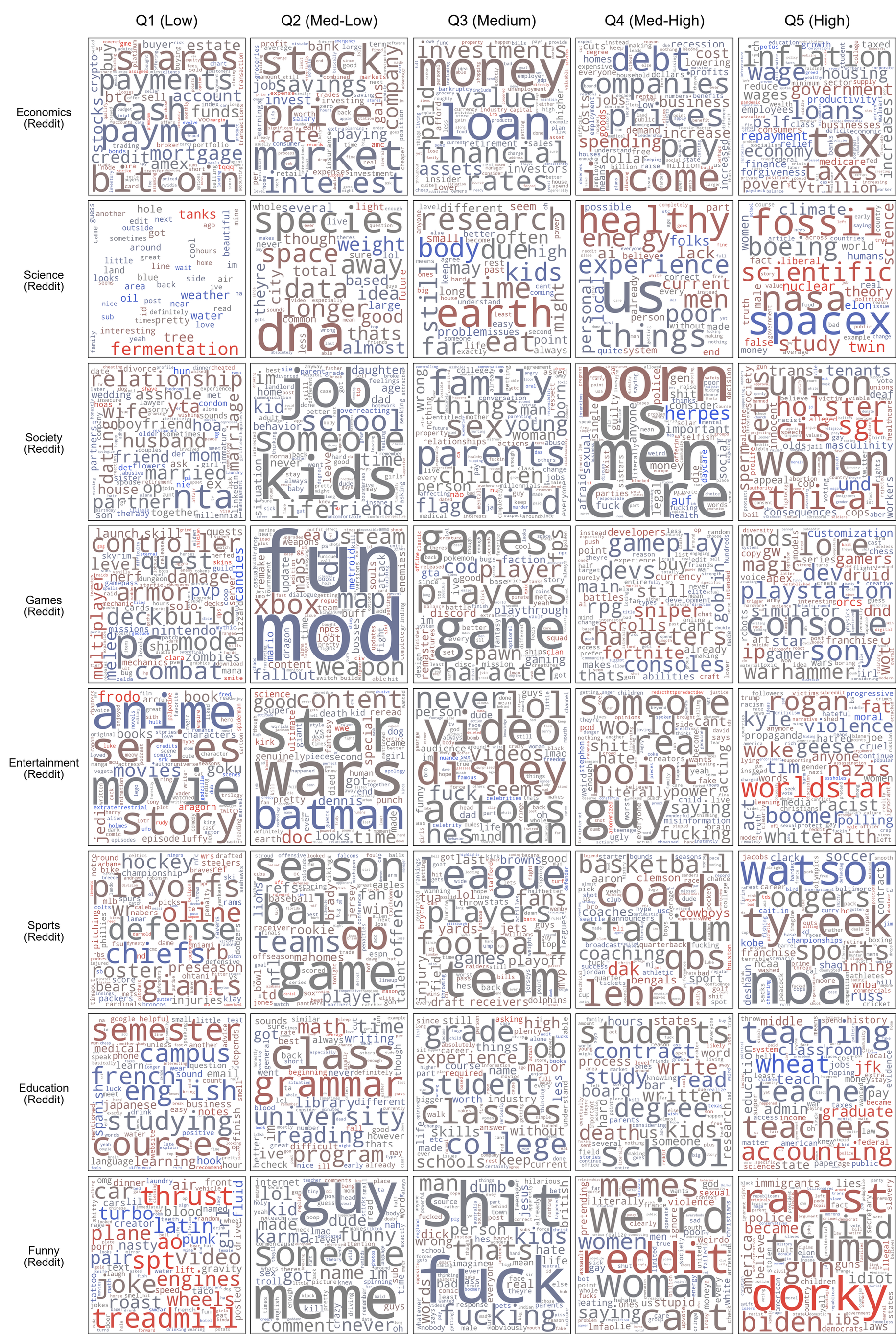}
\caption{Reddit word clouds (Economics, Science, Society, Games, Entertainment, Sports, Education, Funny)}
\label{fig:siwc_reddit2}
\end{figure}

\begin{figure}[h]
\centering
\includegraphics[width=0.8\textwidth]{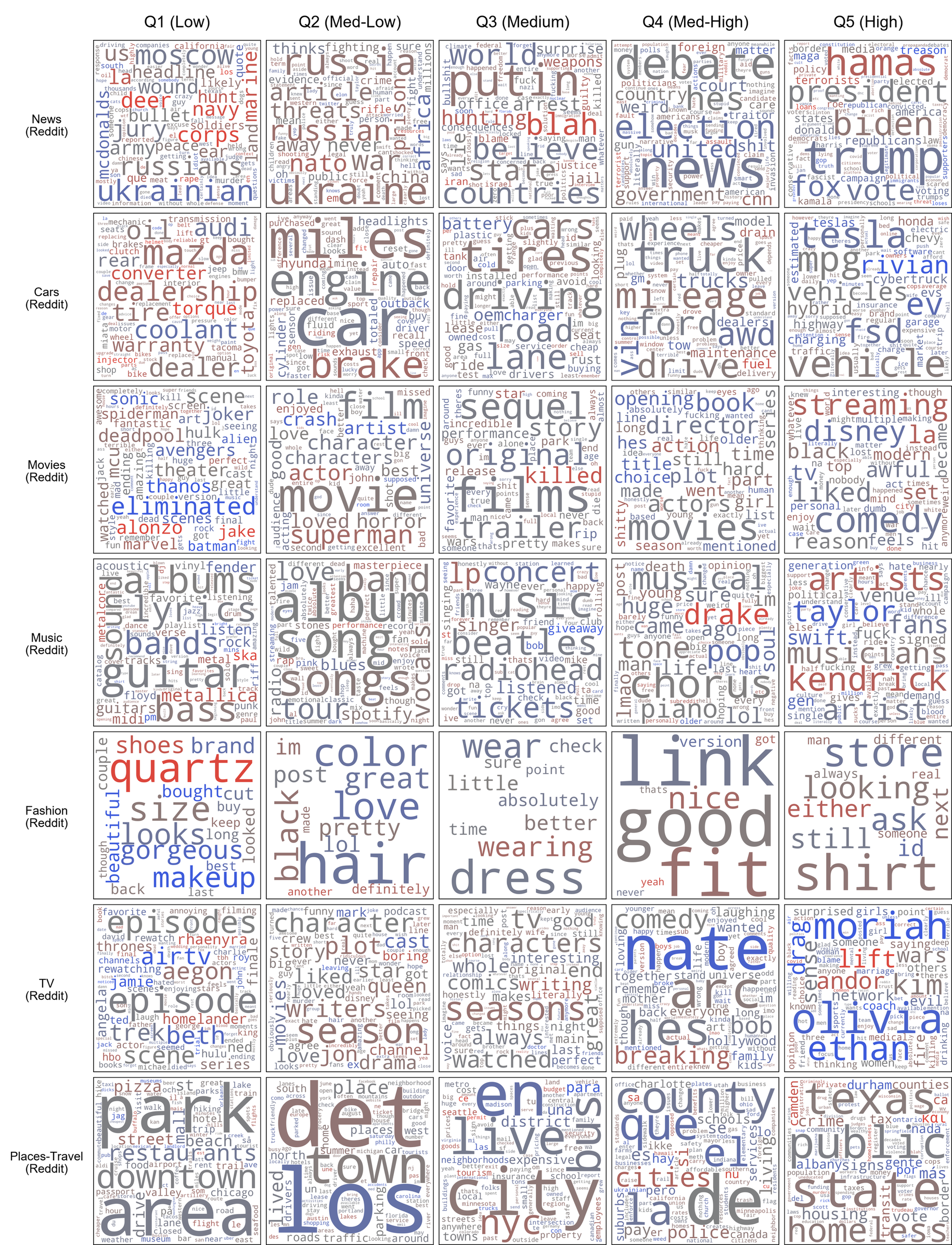}
\caption{Reddit word clouds (News, Cars, Movies, Music, Fashion, TV, Places-Travel)}
\label{fig:siwc_reddit3}
\end{figure}

Figures~\ref{fig:siwc_ddo1}--\ref{fig:siwc_reddit3} display the word clouds for all categories in both DDO and Reddit.

\clearpage

\begin{figure}[h]
\centering
\includegraphics[width=0.8\textwidth]{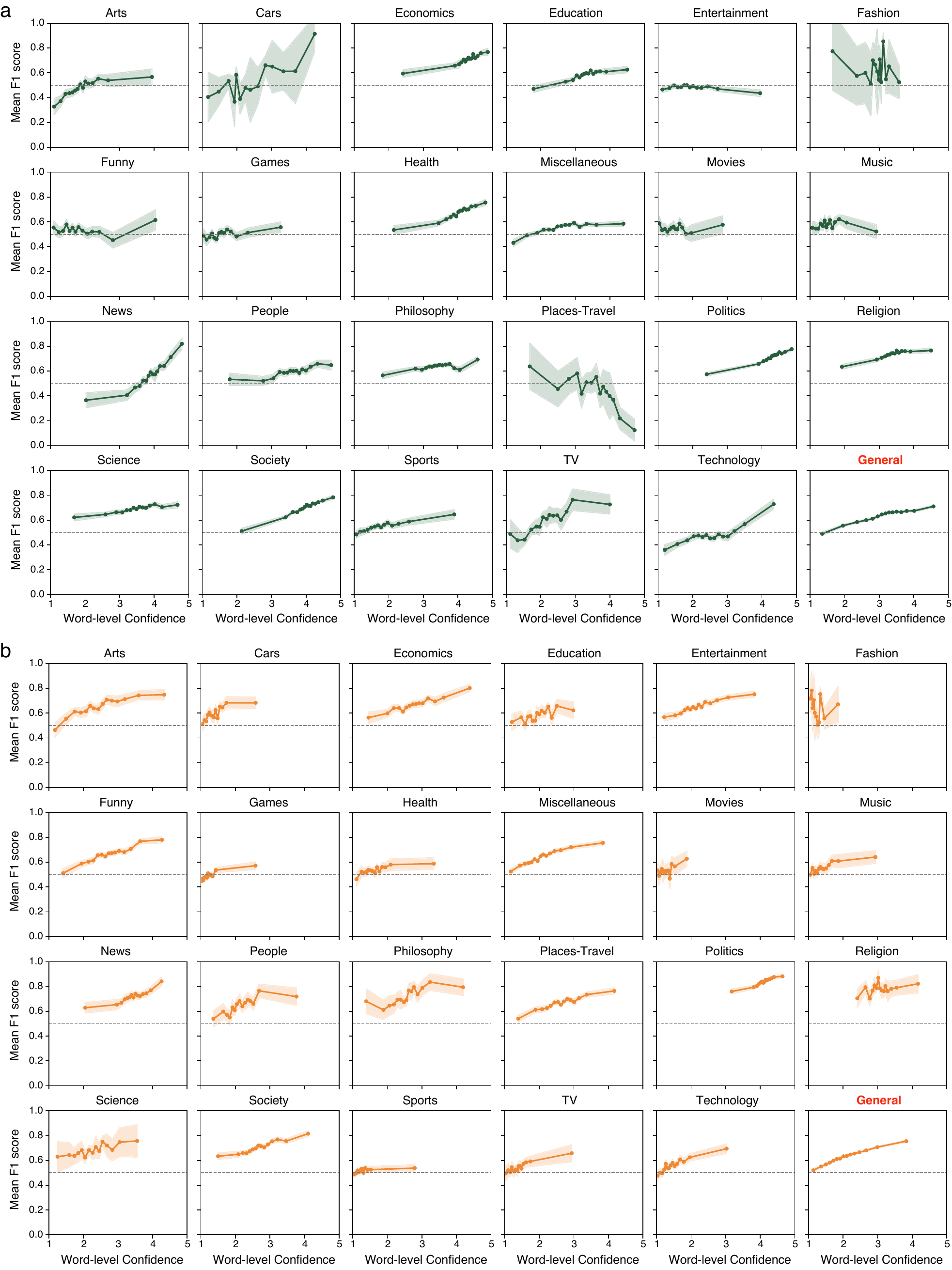}
\caption{\footnotesize Relationship between word-level confidence and mean F1 score of political alignment inference across all categories for (a) the DDO and (b) the Reddit dataset. Each point represents one of 15 confidence quantiles, with shaded areas indicating 95\% confidence intervals. In both (a) and (b), the bottom-right subplot shows the relationship for the entire dataset, which includes all general topic categories.}
\label{fig:conf-f1-multi}
\end{figure}

Figure~\ref{fig:conf-f1-multi} illustrates the relationship between word-level confidence and classification accuracy across categories. 

\clearpage

Table~\ref{tab:answer_ratio} reports on the political orientation inference responses that are formatted correctly across datasets and models.

\begin{table}[h]
\centering
\caption{Number and ratio of correctly formatted political orientation inference responses across datasets and models.}
\begin{tabular}{lrrr}
\hline
Dataset & Total text & \hspace{1em} Answered text & \hspace{1em} Text-level answer ratio \\
\hline
DDO, GPT-4o        & 22,265 & 21,389 & 0.961 \\
DDO, Llama-3.1-8B  & 22,265 & 21,759 & 0.977 \\
Reddit, GPT-4o     & 45,960 & 44,751 & 0.974 \\
Reddit, Llama-3.1-8B & 45,960 & 45,161 & 0.983 \\
\hline
\end{tabular}
\label{tab:answer_ratio}
\end{table}

Table~\ref{tab:maxconf-prop} shows the average numbers of texts per user for the different aggregation methods across datasets and models.

\begin{table}[h]
\centering
\caption{Average number of texts per user contributing to user-level inference across datasets and models.
The table reports: (1) the mean number of texts available for majority or confidence-weighted aggregation (``Total''), (2) the mean number of texts retained under maximum-confidence aggregation, and (3) the proportion of retained texts relative to the total.}
\resizebox{\textwidth}{!}{
\begin{tabular}{llccc}
\hline
\textbf{Dataset} & \textbf{Model} & 
\makecell{\textbf{Mean texts per user}\\\textbf{(Total)}} & 
\makecell{\textbf{Mean texts per user}\\\textbf{(Maximum-confidence)}} & 
\makecell{\textbf{Proportion}} \\
\hline
\multirow{2}{*}{DDO (Combined)} & GPT-4o       & 6.342  & 2.018 & 0.318 \\
                                  & Llama-3.1-8B & 6.240  & 2.987 & 0.479 \\
\hline
\multirow{2}{*}{DDO (Political)}        & GPT-4o       & 3.291  & 1.847 & 0.561 \\
                                  & Llama-3.1-8B & 3.250  & 2.348 & 0.722 \\
\hline
\multirow{2}{*}{DDO (General)}    & GPT-4o       & 5.437  & 1.609 & 0.296 \\
                                  & Llama-3.1-8B & 5.342  & 2.257 & 0.422 \\
\hline
\multirow{2}{*}{Reddit (Combined)} & GPT-4o       & 23.064 & 2.435 & 0.106 \\
                                     & Llama-3.1-8B & 22.931 & 3.920 & 0.171 \\
\hline
\multirow{2}{*}{Reddit (Political)}     & GPT-4o       & 2.479  & 1.380 & 0.557 \\
                                  & Llama-3.1-8B & 2.471  & 1.574 & 0.637 \\
\hline
\multirow{2}{*}{Reddit (General)} & GPT-4o       & 21.116 & 2.413 & 0.114 \\
                                  & Llama-3.1-8B & 20.990 & 3.190 & 0.152 \\
\hline
\end{tabular}
}
\label{tab:maxconf-prop}
\end{table}

\end{document}